\documentclass[showpacs,showkeys,preprintnumbers,fleqn]{revtex4}
\usepackage{amsmath,amssymb}
\usepackage{graphicx}
\def\bra{\langle}
\def\ket{\rangle}

\def\Kbar{\overline{{\rm K}}}

\newcommand{\FRAC}[2]{\leavevmode\kern.1em
  \raise.5ex\hbox{\the\scriptfont0 #1}\kern-.1em
  /\kern-.15em\lower.25ex\hbox{\the\scriptfont0 #2}}
\newcommand{\xbld}[1]{\mbox{\boldmath $ #1 $}}
\newcommand{\dd}{{\rm d}}
\newcommand{\ii}{{\rm i}}
\begin{document}
\setlength{\baselineskip}{18pt}

\title{$\Lambda$(1405) in a baryon-meson scattering with a bound state embedded \\ in the continuum}
\author{Sachiko Takeuchi}\affiliation{%
Japan College of Social Work, Kiyose, Tokyo 204-8555, Japan}
\author{Kiyotaka Shimizu}\affiliation{%
Department of Physics, Sophia University, Chiyoda-ku, Tokyo 102-8554, Japan
}
\date{\today}


\begin{abstract}
We investigate $\Lambda$(1405) as a resonance in a coupled channel 
baryon-meson ($\Sigma \pi$-N$\Kbar$-$\Lambda \eta$) scattering with a
`bound state 
embedded in the continuum' (BSEC). This BSEC is introduced by hand,
as a state not originated from simple baryon-meson systems.
We assume it comes from the three-quark state.
For this purpose, we solve the Lippmann-Schwinger equation with semirelativistic kinematics
in the momentum space. 

There appears a resonance in the $\Sigma\pi$ scattering below the N$\Kbar$ threshold
when the N$\Kbar$ channel is taken to be strongly attractive.
It occurs without introducing a BSEC, just like the chiral unitary approach.
When a BSEC is introduced, a resonance also appears around at 1405MeV
with a weaker baryon-meson interaction.
The corresponding peak also has a large width, and the N$\Kbar$ scattering length 
is reproduced well.
The interaction whose channel dependence is the same 
as the one originated from the color-magnetic interaction,
where no N$\Kbar$ attraction exists,
also gives a broad peak like $\Lambda$(1405) with help of a BSEC.

It is found that the energy-dependent potential gives a broader peak. 
The BSEC coupling which does not vanish at the zero momentum transfer also give a larger width.
In order to reproduce the observed N$\Kbar$ scattering length, introducing a BSEC seems preferable.
In our calculation, a model gives an appropriate value when
the BSEC contribution to the resonance is roughly half of that of the N$\Kbar$ in size.
\end{abstract}

\maketitle

\section{Introduction}
$\Lambda$(1405), the lowest negative-parity baryon in spite of its non-zero strangeness,
has been investigated for a long time.
Recently, a few works to describe $\Lambda$(1405) as 
a baryon-meson resonance have been reported \cite{Kaiser:1995eg,Oset:1997it,Jido:2003cb,Yamazaki:2002uh}.
This picture 
is interesting because coupling to the mesons is surely important for this broad resonance.
Also, it gives new
 lights on the properties of the $\Kbar$-nuclei 
 \cite{Yamazaki:2002uh,Akaishi:2002bg,Lutz:1997wt,Mares:2006vk,Magas:2006fn}.

One of such works is the chiral unitary approach, which employs 
the lowest-order baryon-meson vertices of the nonlinear chiral Lagrangian
as the baryon-meson interaction in the model\cite{Kaiser:1995eg,Oset:1997it}. 
The Lagrangian of this approach is an flavor-SU(3) extended version; 
the baryon-meson interaction is a flavor-flavor type. In such a case,
 the flavor-singlet state gains a strong attraction.
This attraction,
especially the one in the N$\Kbar$ channel, 
plays a major role to produce the $\Lambda$(1405) resonance. 

Recently, it is argued that
there is no need to include an extra pole
to reproduce the $\Lambda$(1405) peak and the low-energy $\Kbar$N scattering data
when one employs the chiral unitary approach
\cite{Kaiser:1995eg,Hyodo:2008xr}.
Namely, the observables can be explained within this baryon-meson framework without
introducing a pole which is not originated from the baryon-meson interaction.
As we will discuss later, the mechanism is as follows.
Due to the strong attraction in the N$\Kbar$ channel,
there is a N$\Kbar$ bound state in this framework.
When the N$\Kbar$-$\Sigma\pi$ coupling is switched on, this bound state 
becomes a resonance.
While the baryon-meson potential in the $\Sigma\pi$ channel is more attractive than
that of N$\Kbar$ in this approach,
the former channel is not affected much by this attraction.
It is because 
the pion cannot stay at the short distance, where the attraction exists,
due to its light mass.
Thus the short-range part of the resonance wave function is mostly N$\Kbar$'s,
and the resonance stays 
just below the N$\Kbar$ threshold.
In contrast to this $\Lambda$(1405) case,
it seems necessary to include an extra pole to 
produce N(1535) in the chiral unitary approach
\cite{Kaiser:1995cy,Nacher:1999vg,Hyodo:2008xr}.
It is probably reflecting the fact that
the closest attractive channel is $\Sigma$K, which is by 150 MeV above the N(1535) peak.
Thus, the mechanism which works well for $\Lambda$(1405) does not work for N(1535).

The above situation, however, raises a new question.
The ground state baryons are described quite nicely by the three-quark (q$^3$) states
in the constituent quark model.
If one extends this idea to the negative parity baryons, there should be 
 a flavor-singlet orbitally-excited q$^3$ state near $\Lambda$(1405) \cite{Isgur:1978xj,FST03}.
This state is supposed to be seen 
as another peak associated with the same quantum number,
or, at least, affects the baryon-meson scattering largely.
Thus, introducing such a pole will
destroy the above baryon-meson picture for $\Lambda$(1405).
The theoretical reason 
why an extra pole should be introduced into N(1535) 
but not into $\Lambda$(1405) is not clear yet.

Moreover, 
it has been shown that no large N$\Kbar$  attraction appears  
if one construct
the baryon-meson interaction 
from the gluonic interaction between quarks in the quark model. 
 It is, however, possible to reproduce the $\Lambda$(1405) resonance successfully.
There, the flavor-singlet q$^3$ state plays an important role 
to reproduce the peak \cite{Takeuchi:2007tv}.
Thus, there are still ambiguities in choosing the baryon-meson interaction.

In this paper, we compare the models  
and clarify the mechanism to form the $\Lambda$(1405) peak.
For this purpose,  we solve 
the Lippmann-Schwinger equation with semirelativistic kinematics in the momentum space.
We take following two baryon-meson interactions,
which have different channel dependence from each other:
one is the flavor-flavor type interaction (FF-type) and the other is that
obtained from the color-magnetic interaction (CMI) of the quark model (CMI-type).
The orbital part of the interaction we employ here is a simple separable one 
with the gaussian form factor for all the cases for the sake of simplicity.  
By removing the difference in the orbital part, we can concentrate on the effects from the
channel dependence and the strength of the interactions.

In addition to the above usual baryon-meson channels,  
we also introduce a `bound state 
embedded in the continuum' (BSEC). 
Here we consider that this BSEC corresponds to the flavor-singlet q$^3$ state predicted 
by quark models.
As for the transition potential from the baryon-meson system
to the negative parity baryon,
we employ the constituent quark model to analyze the 
structure of the vertex B($\frac{1}{2}^+$)B($\frac{1}{2}^-$)M($0^-$).  
Since this BSEC is flavor-singlet, for the FF-type model,
the transition potential from a baryon-meson channel
is proportional to the size of the flavor-singlet component in that channel.
As for the CMI-type, we use the one which comes from the 
pair-annihilation diagram with the one-gluon exchange in the quark model \cite{Takeuchi:2007tv}.
By adjusting the gaussian cut-off parameter and the size of the BSEC coupling,
we fit the resonance energy.

Note that, since the quark degrees of freedom are not taken
into account directly, our calculation with the CMI-type interaction differs from the
original quark-model calculation.  Nevertheless we use this simplified calculation to discuss the quark picture 
because the quark Pauli-blocking effects are small in this channel.  
In such a case, it has been found that
the results from the quark model can be expressed approximately
by an energy-independent baryon-meson interaction \cite{Takeuchi:2002cw}.
In fact, the obtained results in the present work are similar to the original ones.
We also would like to mention that using the CMI-type interaction in this way has 
an advantage: the system can be treated semirelativistically, which is
difficult within the usual quark-cluster-model framework.

As we will show later, it is found that 
the chiral unitary approach is not a unique way to produce the $\Lambda$(1405) resonance.
With help of BSEC, a weaker FF-type interaction or the CMI-type interaction 
can also give an appropriate peak.
To clarify the difference of their mechanisms,
we investigate the relative importance of roles of BSEC and of the baryon-meson attractions.
To see this, we calculate the probability of the 
BSEC at the resonance energy and compare that with the probabilities 
of the baryon-meson states of the closed channels.
The wave function in the coordinate space, which we use to calculate the baryon-meson probabilities, 
is obtained by 
the Fourier transformation from the off-shell T-matrix.  

Resent experiments have discovered many hadron resonances which may have multiquark components
\cite{pdg06}.
The present method to handle a BSEC and scattering states simultaneously
may be applied to various such exotic systems.

In the next section, we explain the SU(3) structure of the baryon-meson system and the 
baryon-meson interactions. Then, in section II B, we explain how to solve the scattering problem 
with a BSEC in the  
coupled channel baryon-meson scattering problems.
The Lippmann-Schwinger equation is solved  in the momentum space with
the semirelativistic approach. 
 The resonance wave function in the coordinate space is derived in section II C.
In section III, we show the obtained phase shifts and
the mass spectrum of the $\Sigma \pi$ channel as well as the resonance wave functions. 
Discussion on the difference of the mechanisms to reproduce $\Lambda$(1405) 
among the above models is also given.
Summary is given in section IV.   

\section{Model}
%%%
%%%
\subsection{Model space and the interactions} 
%%%
%%%
The flavor-octet baryon and meson system can be classified into six
representations:
\begin{equation}
\xbld{8}_B \times \xbld{8}_M = \xbld{1}_{BM}+\xbld{8}^A_{BM}+\xbld{8}^S_{BM}+\xbld{10}_{BM}+
\overline{\xbld{10}}_{BM}+\xbld{27}_{BM} .
\end{equation}
The strangeness=$-1$ and isospin $T=0$ state appears in the 
$\xbld{1}_{BM}$, $\xbld{8}^A_{BM}$, $\xbld{8}^S_{BM}$, and $\xbld{27}_{BM}$ states.
These four states are given by a linear 
combination of the following four baryon-meson systems:
$
\Sigma \pi,\hspace{5pt}{\rm N} \Kbar,\hspace{5pt} \Lambda \eta,\hspace{5pt} \Xi {\rm K}
$.
Since the threshold energy is much higher, 
the $\Xi$K channel does not affect the $\Lambda$(1405) peak much 
as we will show later.

The hamiltonian is divided into the 
baryon-meson space (the $P$-space) and the 
q$^3$ pole space (the $Q$-space):
\begin{equation}
H = \left( \begin{matrix} H_P & V_{PQ} \\ V_{QP} & E_Q\end{matrix} \right) ,
\end{equation}
and the wave function is given by
\begin{equation}
\psi=\left( \begin{matrix} \psi_P \\
\psi_Q\end{matrix} \right) .
\end{equation}

Now we explain the interactions between baryon and meson,  $V_{P}\equiv V_{BM}$,
 and the coupling of the 
baryon-meson state with the BSEC such as the q$^3$ state, $V_{QP}$.

The potential between baryon and meson is assumed to be a central separable potential.
The one between the $i$-th and $j$-th baryon-meson channels is defined by
\begin{equation}
V_{ij}(\xbld{p},\xbld{p}')=c_{ij} {\pi V_0\over 2}u 
\exp [-\frac{1}{4}{a^2}(p^2+p'^2)]Y_{00}(\Omega_p)Y_{00}(\Omega_{p'})
=c_{ij} {V_0\over 8}u 
\exp [-\frac{1}{4}{a^2}(p^2+p'^2)] .
\end{equation}
Here, $V_0$ and $a$ are strength and range of the  potential.  
The factor $u$ is taken to be 1 for an energy-independent potential while
for an `energy-dependent' potential 
\begin{equation}
u=\frac{\sqrt{k^2+m^2}}{m}~,
\end{equation}
where $m$ and $k$ are meson mass and initial momentum, respectively.
This factor is introduced to take the
energy dependence like the
WT term into account.

Note that the potential acts only on the S-wave baryon-meson states. 
For the FF type, 
the factor $c_{ij}$ is taken to be
\begin{equation}
c_{ij}=(\xbld{F}_{B} \cdot \xbld{F}_{M})_{ij} ,
\end{equation}
where
$\xbld{F}_{B}$ and $\xbld{F}_{M}$ are SU(3) flavor generators for baryon and meson.
In the CM type, $c_{ij}$'s are given by the color-magnetic interaction with 
the the quark exchanges.
The matrix elements of $c_{ij}$ for both of the cases for strangeness=$-1$ and isospin $T$=0 are shown in Tables I and II.

\begin{table*}[tb]
\caption{The factor $c_{ij}$ for the FF type}
\vspace{5mm}
\renewcommand\arraystretch{1.8}
\begin{tabular}{|c|c|c|c|c|} \hline
 & $\Sigma \pi$ & N$\Kbar$  & $\Lambda \eta$ & $\Xi K$ \\ \hline
 $\Sigma \pi$ & -8 & $\sqrt{6}$ &
 0 & -$\sqrt{6}$ \\ \hline
 N$\Kbar$  & & -6 & $3\sqrt{2}$ & 0   \\ \hline
 $\Lambda \eta$ & & & 0 & -$3\sqrt{2}$ \\ \hline
$\Xi {\rm K}$  & & & & -6 \\ \hline
$\xbld{1}_{BM}$  &$\sqrt{3\over 8}$ &$-{1\over 2}$ & $\sqrt{1\over 8}$ & ${1\over 2}$ \\ \hline
\end{tabular}
\end{table*}

\begin{table*}[tb]
\caption{The factor $c_{ij}$ for the CM type}
\vspace{5mm}
\renewcommand\arraystretch{1.8}
\begin{tabular}{|c|c|c|c|c|} \hline
 & $\Sigma \pi$ &  N$\Kbar$  & $\Lambda \eta$ & $\Xi {\rm K}$ \\ \hline
 $\Sigma \pi$ & $-{16\over 3}$ & ${116\sqrt{7}\over 21}$ &  $-{16\sqrt{105}\over 105}$ & 0 \\ \hline
 N$\Kbar$     & & 0 & ${28\sqrt{15}\over 15}$ & 0   \\ \hline
 $\Lambda \eta$ & & & ${112\over 15}$ & $-{40\sqrt{70}\over 21}$ \\ \hline
$\Xi$K  & & & &$-{160\over 21}$ \\ \hline
q$^3$  &140 &$-85$ & 53 & -\\ \hline
\end{tabular}
\end{table*}

The coupling of the baryon-meson state with the $Q$ state $\bra Q|V|\xbld{p}\ket$ is given by 
%$<Q | V | R_j>$ which is %evaluated by taking a 
the following gaussian form,
\begin{equation}
\bra Q|V|\xbld{p}\ket=V_0^{QP} %(\sqrt{\pi} a_Q)^{-3/2} (a_Q)^3 (a_Q p)^2 
\{ c_1+c_2(a_Q p)^2 \} \exp[-\frac{1}{4}{a_Q^2}p^2]\sqrt{4\pi}Y_{00}(\Omega_p) , \label{vpq}
\end{equation}
where $V_0^{QP}$ is the strength and $a_Q$ describes the form factor of the coupling potential.
We take the $c_1+c_2(a_Q p)^2$ dependence on the $p^2$ for the transition potential 
from the relative S-wave baryon-meson state to the negative parity baryon state $Q$. The origin of the form of this 
transition potential is explained in Appendix B. 

As for the $Q$ state, namely, q$^3$ system, which is treated as BSEC, 
we assume the state to be a flavor singlet state for the FF type. 
The flavor singlet baryon meson 
state  $|\xbld{1}_{BM}\ket$ is given by 
\begin{equation}
|\xbld{1}_{BM}\ket= \sqrt{\frac{3}{8}}|\Sigma \pi\ket-\frac{1}{2}|{\rm N} \Kbar \ket+
\sqrt{\frac{1}{8}}|\Lambda \eta\ket+
\frac{1}{2}|\Xi {\rm K}\ket .
\end{equation}
This fixes the relative strength of the coupling potential of the FF type with the flavor singlet $Q$ state 
among the baryon-meson 
channels.

\subsection{Lippmann-Schwinger equation with BSEC}

The Lippmann-Schwinger equation for $H=H_0+V$ is written 
\begin{equation}
T=V+VG^{(0)}T, \hspace{10pt}G^{(0)}=\frac{1}{E-H_0+{\rm i}\varepsilon} .
\end{equation}
%We divide a space into $P$ (baryon-meson space) and $Q$ (BSEC space). 
We assume that the $Q$ space has only one state, then we can set
$QHQ \equiv E_Q$ or
$QVQ \equiv V_{QQ}=0$.
Using $P+Q=1$, 
we obtain the $T$ matrix of the $P$ space, $T_{PP}$, as the following form \cite{Feshbach}.
For details, see appendix A.
\begin{equation}
T_{PP}=T^{(P)}+(1+V_{PP}G_P)V_{PQ}G_QV_{QP}(1+G_PV_{PP}) \label{tpp} .
\end{equation}
Here the first term on the right hand side is the $T$-matrix solved 
within the $P$ space:
\begin{equation}
T^{(P)}=(1-V_{PP}G^{(0)}_P)^{-1}V_{PP}.
\end{equation}
The $G_{P}$ is a propagator in $P$ space which is given by  
\begin{equation}
G_{P}=G^{(0)}_P(1-V_{PP}G^{(0)}_P)^{-1}=(G^{(0)-1}_P-V_{PP})^{-1},
\end{equation}
and $G_{Q}$ is a propagator for the $Q$ space, which contains the coupling with the $P$ space:
\begin{equation}
G_{Q}=G^{(0)}_Q(1-V_{QP}G_{P}V_{PQ}G^{(0)}_Q)^{-1}=(G^{(0)-1}_Q-V_{QP}G_{P}V_{PQ})^{-1}.
\end{equation}
The term $(1+V_{PP}G_P)$ describes a distortion in the $P$ space due to the potential $V_{PP}$.

Because we consider light meson systems, we take the semirelativistic kinematics for the 
baryon and meson propagators.
The free propagator for the initial energy $E_{tot}$ % ($k^0,-\xbld{p}$) and ($p^0,\xbld{p}$) 
in the $P$ space becomes
\begin{eqnarray}
G_P^{(0)}&=& 
\ii \int {d^4q\over(2\pi)^4} \;
{M\over \Omega} \;
{1\over E_{tot}%k^0 + p^0
-q^0 -\Omega+\ii \varepsilon} \;
{2m\over q_0^2 -\xbld{q}^2-m^2 +\ii \varepsilon}
\\
&=&
\int {d^3 \xbld{q}\over(2\pi)^3} \;
{M\over \Omega} \;
{1\over E_{tot}%k^0 + p^0
-\omega -\Omega+\ii \varepsilon} \;
{m\over \omega} \label{gfunc} ,
%\int k^2dk \;
\end{eqnarray}
where
\begin{equation}
\Omega = \sqrt{M^2+\xbld{q}^2},~~~ \omega=\sqrt{m^2+\xbld{q}^2}~~~\text{and}~~~
E_{tot}=\sqrt{M^2+\xbld{k}^2}+\sqrt{m^2+\xbld{k}^2} . 
\end{equation}
Here the $\xbld{k}$ is the relative momentum of the baryon-meson system and $M$ and $m$ 
are the baryon and meson masses, respectively. We include a factor $2m$ in the propagator so that 
the propagator becomes the usual one in the non-relativistic limit.
The factor ${mM\over \omega \Omega}$ can be taken into account in the potential, as 
a kind of form factor.
\begin{equation}
V_{ij}(\xbld{p},\xbld{p}') \rightarrow \sqrt{mM\over \omega \Omega}
V_{ij}(\xbld{p},\xbld{p}')\sqrt{m'M'\over \omega' \Omega'} .
\end{equation}
Note that this factor plays a very important role to reduce the strength of the potential 
in a high momentum region. The effect is strong when the mass is small. Therefore this factor strongly cuts off the 
$\Sigma \pi$ potential.  
In the actual calculation, the factor ${mM\over \omega \Omega}$ is 
taken into account in a factor $C$ which takes into account a difference between 
relativistic and non-relativistic kinematics.

Let us introduce the factor $C$ in the followings.
The hamiltonian of the $P$ space, $H_P$, becomes:
\begin{eqnarray}
H_P=H_0+V_P=\Omega+\omega+V_P .
%\sqrt{p^2+M^2}+\sqrt{p^2+m^2}
\end{eqnarray}
For the Green function $G^{(0)}$, we use
\begin{equation}
(E_{tot}-H_0){\omega \Omega \over Mm}  = {k^2-q^2\over 2\, \mu \,C(q,k)} ,
\end{equation}
where $\mu$ is the reduced mass,
$\mu = {Mm\over M+m}$,
and $C$ is a positive function of $p$ and $k$.
The variable $k$ is the on-shell momentum while $q$ is the momentum operator.
The factor $C$ is 1 for the nonrelativistic system. For 
the semirelativistic case, where $E_{tot}-H_0$ is given by 
\begin{equation}
E_{tot}-H_0 = \sqrt{M^2+k^2}+\sqrt{m^2+k^2}
-\sqrt{M^2+q^2}-\sqrt{m^2+q^2} ,
%\sqrt{\kappax^2+m^2_1}+\sqrt{\kappax^2+m^2_2}-\sqrt{\kx^2+m^2_1}-\sqrt{\kx^2+m^2_2} ,
\end{equation}
 the factor $C$ becomes
\begin{equation}
C(q,k)={q^2-k^2\over
2\mu(\sqrt{M^2+k^2}+\sqrt{m^2+k^2}
-\sqrt{M^2+q^2}-\sqrt{m^2+q^2})}{mM\over \omega \Omega}.
\end{equation}
Note that the factor $C(q,k)$ is positive for any real $q$ and $k$.
 
%The condition $C(k,k_0)>0$ holds because
%\begin{equation}
%C(k_0,k_0)={\sqrt{k^2_0+m^2_1}\sqrt{k^2_0+m^2_2}
%\over
%m(\sqrt{k^2_0+m^2_1}+\sqrt{k^2_0+m^2_2})}.
%\end{equation}

When we use 
\begin{equation}
{\tilde V}=\pi \mu V ~~~\text{and}~~~{\tilde T}=\pi \mu T,
\end{equation}
then the Lippmann-Schwinger equation becomes
\begin{equation}
{\tilde T} = {\tilde V}+{2\over \pi} \int {\tilde V}|q\ket
{C(q,k) q^2 {\rm d}q\over q^2-k^2-\ii\varepsilon}\bra q|{\tilde T}. \label{lseqn}
\end{equation}
The $S$ matrix can be obtained from this ${\tilde T}$ as
\begin{equation}
S = 1- 2\ii\,k \,C(k,k)\, {\tilde T}.
\end{equation}
The reason why we have introduced the factor $C(q,k)$ is the following. 
%The main reason is due to a technical procedure to solve the eq. (\ref{lseqn}). 
The propagator in eq.\ (\ref{gfunc}) has a pole at $q=k$. 
 By introducing the factor $C$, one can 
 separate the propagator into 
a simple form $1/(q^2-k^2)$ and a smooth function $C(k,q)$
instead of dealing with the complicated singular function of $q^2$ directly. 
When we solve the Lippmann-Schwinger 
equation 
numerically, we simply take care of the pole of the $1/(q^2-k^2)$
by the well known procedure.

\subsection{Wave functions}

The wave function $\psi$ with the initial 
wave function $\phi_{ini}$ can be obtained by using the $T$-matrix:
\begin{eqnarray}
|\psi\ket &=&|\phi_{ini}\ket +G^{(0)} T |\phi_{ini}\ket .
\end{eqnarray}
The wave function of the $P$-space is
\begin{eqnarray}
|\psi_P\ket &=&|\phi_{ini}\ket +G^{(0)}_P T_{PP} |\phi_{ini}\ket ,
\end{eqnarray}
whereas the wave function for the $Q$-space becomes
\begin{eqnarray}
|\psi_Q \ket&=& G^{(0)}_Q T_{QP} |\phi_{ini}\ket .
\end{eqnarray}

Relative importance between the closed $P$-space $\phi_c$ (different from $\phi_{ini}$) and the $Q$-space 
can be found by comparing the following probabilities.
\begin{eqnarray}
|\bra \phi_c |\psi_P\ket|^2 &=&|\bra| \phi_c|G^{(0)}_P T_{PP} |\phi_{ini}\ket|^2 , 
\label{eq:28}\\ 
|\bra Q|\psi_Q\ket|^2 &=&|\bra Q|G^{(0)}_Q T_{QP} |\phi_{ini}\ket|^2 .\label{eq:29}
%=|G_QV_{QP}(1+G_PV_{PP})\phi_{ini}|^2
%=|\bra Q|G_QV_{QP}(M^t)^{-1}|\phi_{ini}\ket|^2.
\end{eqnarray}
The wave function of the coordinate space can be obtained by the 
Fourier transformation.
In the following calculation, the asymptotic wave function of 
 the $\Sigma \pi$ channel is normalized as $2 \sin (kr + \delta)/k$.

\section{Results}

Here we show the obtained phase shift $\delta$ 
of the  $\Sigma\pi$ 
channel for the relative 
angular momentum $L=0$. 
We also show the $\Sigma\pi$ mass spectrum given by
\begin{equation}
|1-\eta {\rm e}^{2{\rm i}\delta}|^2/k ~,
\end{equation}
where $\eta$ is the elasticity and $k$ is the 
relative wave number for the $\Sigma\pi$ scattering. 
This mass spectrum corresponds to the observed resonance peak.

Here we take the model parameters to adjust the peak properties to the experimental values:
 the resonance energy is
1406 $\pm$ 4 MeV  and the width is 50 $\pm$ 2 MeV \cite{pdg}.
There is an argument that the $\Lambda$(1405) has a two-pole structure,
and that the peak energy found in the $\Sigma\pi$ scattering and that in the N$\Kbar \rightarrow \Sigma\pi$ decay 
may be different \cite{Magas:2005vu,Jido:2005ew}.
Both of the two peak energies, 1390 and 1426 MeV, however, are far above the 
$\Sigma\pi$ threshold and below the N$\Kbar$ threshold.
In that sense, our conclusion will not change if we use the results of their analysis. 

We use the experimental values for the masses of baryons and mesons in the kinematics,
which are shown in 
Table III. 
In the present work, we consider eight parameter sets, which are listed in Table IV. 
The obtained mass spectra and the phase shifts as well as the wave function at the resonance
 for each of the parameter sets are shown in Figs.\ 1-9.
In Table V, we summarize the obtained resonance energy  and
the N$\Kbar$ scattering length.
 The observed N$\Kbar$ scattering length is 
$(-1.70 \pm 0.07) + (0.68 \pm 0.04)\,\ii$ fm \cite{Martin:1980qe}.
The probabilities of the closed channels and the q$^3$ state,
which are given by eqs.\ (\ref{eq:28}) and (\ref{eq:29}), are also shown in Table V.
These q$^3$ probabilities indicate the relative importance among these components.
They should not be compared with those from the different parameter set
because it is
the wave function of the open channel that is normalized.
The relative importance of the closed channels to the open channel 
 can be seen by comparing the short-range part of the resonance wave functions. 
The self energies gained by the BSEC for the parameter set (4)-(8) are shown in Table VI.

\begin{table}[btp]
\begin{center}
\caption{Masses of baryons and  mesons in [MeV] \cite{pdg}.} 

\vspace{5mm}

\begin{tabular}{ |cccccccc|} \hline
 & N & $\Sigma$ & $\Lambda$ & $\Xi$ & $\pi$ & $\Kbar$,K & $\eta$   \\ \hline
 & 939 & 1193 &1116 & 1318 &138 & 496 & 549  \\ \hline
\end{tabular}
\end{center}
\end{table}

\begin{table}[btp]
\caption{Parameters of the potentials} 
\begin{center}
\begin{tabular}{|l|cccccccc|} \hline
  & $a$ & $V_0$ & $a_Q$ & $E_Q$ & $V_0^{PQ}$ & $c_1$ & $c_2$ &Energy dep.\\\
  & fm & MeV$\cdot$fm$^3$ & fm & MeV & MeV$\cdot$fm$^{3/2}$ & & & \\ \hline
(1) Oset-type  & 0.38 & 2.7 & - & - & 0   & - &- &yes\\ \hline
(2) FF without pole  & 0.37 & 2.7 & - & - & 0 & - &- &no\\ \hline
(3) FF without pole  & 0.46 & 2.7 & - &  - & 0 & - &- &no\\ \hline
(4) FF with pole  & 0.46 & 2.7 & 0.46 &  200 & 11.5 & 0.0 &1.0 & both\\ \hline
(5) FF with pole  & 0.46 & 2.7 & 0.46 &  200 & 15.0 & 1.0 &0.0 &no\\ \hline
(6) CMI with pole  & 0.40 & 0.6 & 0.40 & 160 & 12.5 & 1.0 & 0.0 &no \\ \hline
(7) CMI with pole  & 0.40 & 0.8 & 0.40 & 160 & 16.0 & 1.0 & 0.0 &no \\ \hline
(8) CMI with pole  & 0.40 & 0.8 & 0.40 & 160 & 12.0 & 0.0 & 1.0 &no \\ \hline
\end{tabular}
\end{center}
\end{table}

\begin{table}[btp]
\caption{Peak energy, $E_{res}$, the probabilities, and the N$\Kbar$ (T=0) scattering length, $a_{N\Kbar}$, for each parameter set. The values for (4) are the energy-independent ones.
} 
\begin{center}
\begin{tabular}{|l|ccccc|} \hline
  & $E_{res}$ &  \multicolumn{3}{c}{Probabilities}  & $a_{N\Kbar}$\\
  & MeV &N$\Kbar$ &  $\Lambda \eta$ & q$^3$   & fm  \\ \hline
(1) Oset-type  & 1406.9 & 34.4 & 1.3 & - & $-$2.09 + 0.59 i \\ \hline
(2) FF without pole  & 1407.9 & 76.6 & 3.0 & - & $-$1.93 + 0.25 i  \\ \hline
(3) FF without pole  & 1432.0 & 337.8 & 3.3 &  - & $-$4.78 + 1.48 i  \\ \hline
(4) FF with pole  & 1404.0 & 60.5 & 3.1 &  ~42.6 & $-$1.09 + 0.18 i  \\ \hline
(5) FF with pole  & 1403.5 & 35.8 & 1.5 &  ~14.8 & $-$1.65 + 0.43 i  \\ \hline
(6) CMI with pole  & 1405.9 & 11.8 & 0.6 & ~32.7 & $-$0.64 +  0.25 i \\ \hline
(7) CMI with pole  & 1406.2 & 7.9 & 0.3 & ~21.3 & $-$0.67 +  0.34 i \\ \hline
(8) CMI with pole  & 1402.8 & 15.3 & 1.7 & 203.3 & $-$0.01 +  0.03 i \\ \hline
QCM\cite{Takeuchi:2007tv}  & 1404~~ & 8.6 &0.6  &~23.6  & $-$0.75 +  0.38 i \\ \hline
\end{tabular}
\end{center}
\end{table}

\begin{table}[btp]
\caption{Self energy of BSEC} 
\begin{center}
\begin{tabular}{|l|c|} \hline
  & $\Sigma$  \\
  & MeV  \\ \hline
(4) FF with pole  & $-119.0 - 17.1\, \ii$  \\ \hline
(5) FF with pole  & $-104.2 - 49.1\, \ii$  \\ \hline
(6) CMI with pole  & $ -82.8-28.7 \,\ii$\\ \hline
(7) CMI with pole  & $-78.4-40.2 \,\ii$ \\ \hline
(8) CMI with pole  & $-87.8-4.4 \,\ii$ \\ \hline
\end{tabular}
\end{center}
\end{table}

The first parameter set in Table IV
is energy-dependent FF-type without BSEC, which we call Oset-type.
The strength of the FF-type potential roughly corresponds to the 
one given by the chiral unitary model as
\begin{equation}
V_0 \sim {1\over (2\pi)^3 f^2},
\end{equation}
which is about 2.7 [MeV fm$^3$]. 
Here we have used $f=1.15 \times f_{\pi}$ where the pion 
decay constant is $f_{\pi}$=93 MeV \cite{Oset:1997it}.   
We have fixed this value for other  
calculations of the FF-type potential 
as seen in the Table IV. 
We employ the gaussian 
 form factor which corresponds to a size of the baryon in the quark model
while 
a sharp cut-off for the integral of the off-shell momentum
is employed in ref.\ \cite{Oset:1997it}. 
In spite of a difference between these 
treatments, the results of this parameter set is very much alike to 
those in ref.\cite{Oset:1997it}.

In Fig.\ 1, we show  the results of the calculation with the three channels, 
 $\Sigma\pi$, N$\Kbar$ and $\Lambda \eta$. 
In order to see the effect of the $\Xi $K channel, 
 we have also carried out the four-channel calculation. 
As seen in the figure, the effect of the 
 $\Xi$K channel is small and we consider only three-channel calculation 
 in the following.
 The wave functions around the resonance region, where the relative momentum of the 
 $\Sigma \pi$ channel is $k=0.76 {\rm fm}^{-1}$ are also shown in the figures. 

\begin{figure}
\caption{Mass spectrum, phase shift and wave functions of $\Sigma \pi$ scattering
by Oset-type, the energy-dependent potential with the parameter set (1) in Table IV.} 

\includegraphics[height=4.0in]{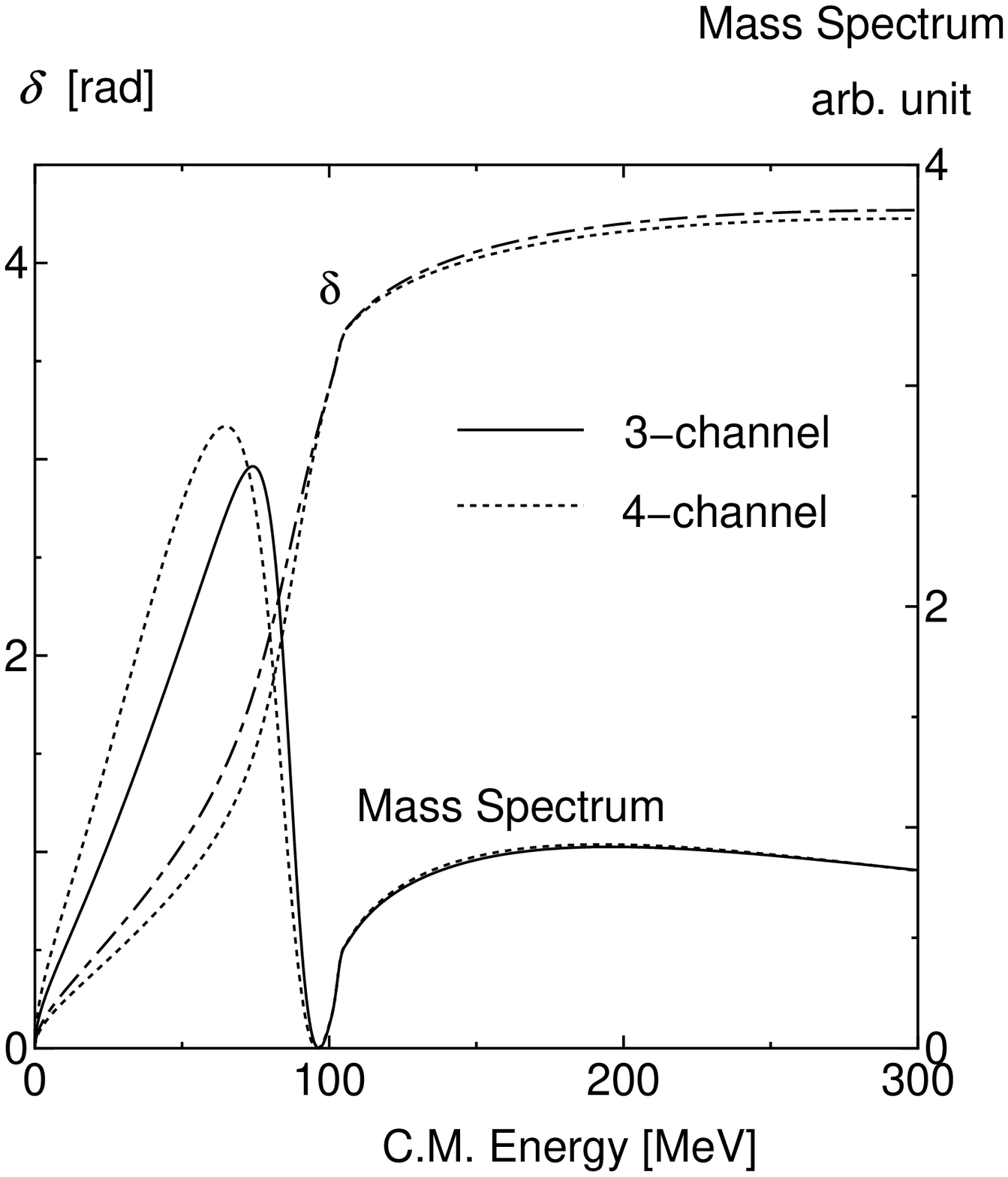}
\includegraphics[height=3.7in]{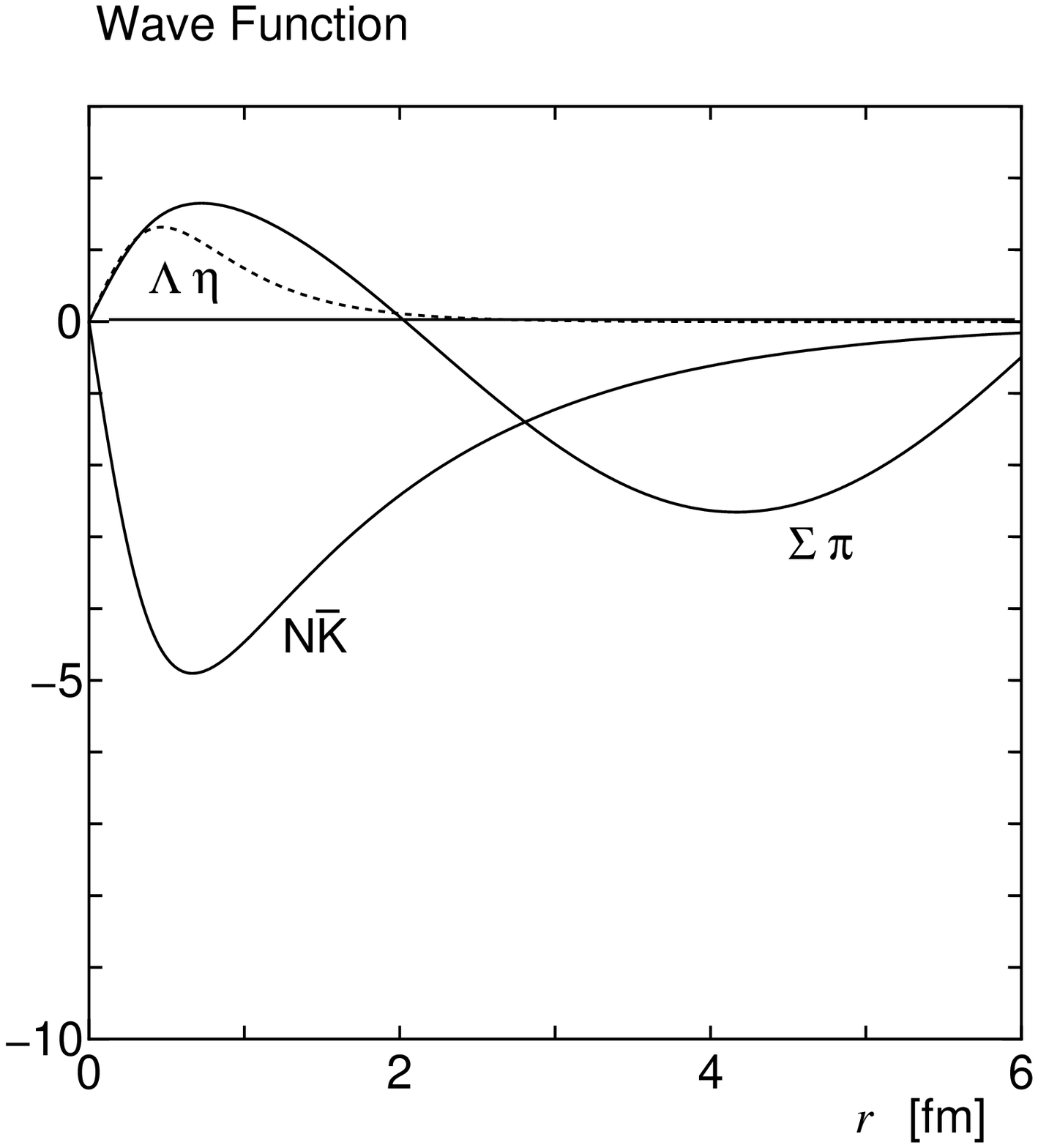}

\label{fg1}
\end{figure}

\begin{figure}
\caption{Mass spectrum, phase shift and wave functions of $\Sigma \pi$ scattering
by the energy-independent FF-type potential with the parameter set (2) in Table IV.
No coupling with the q$^3$ state.} 

\includegraphics[height=4.0in]{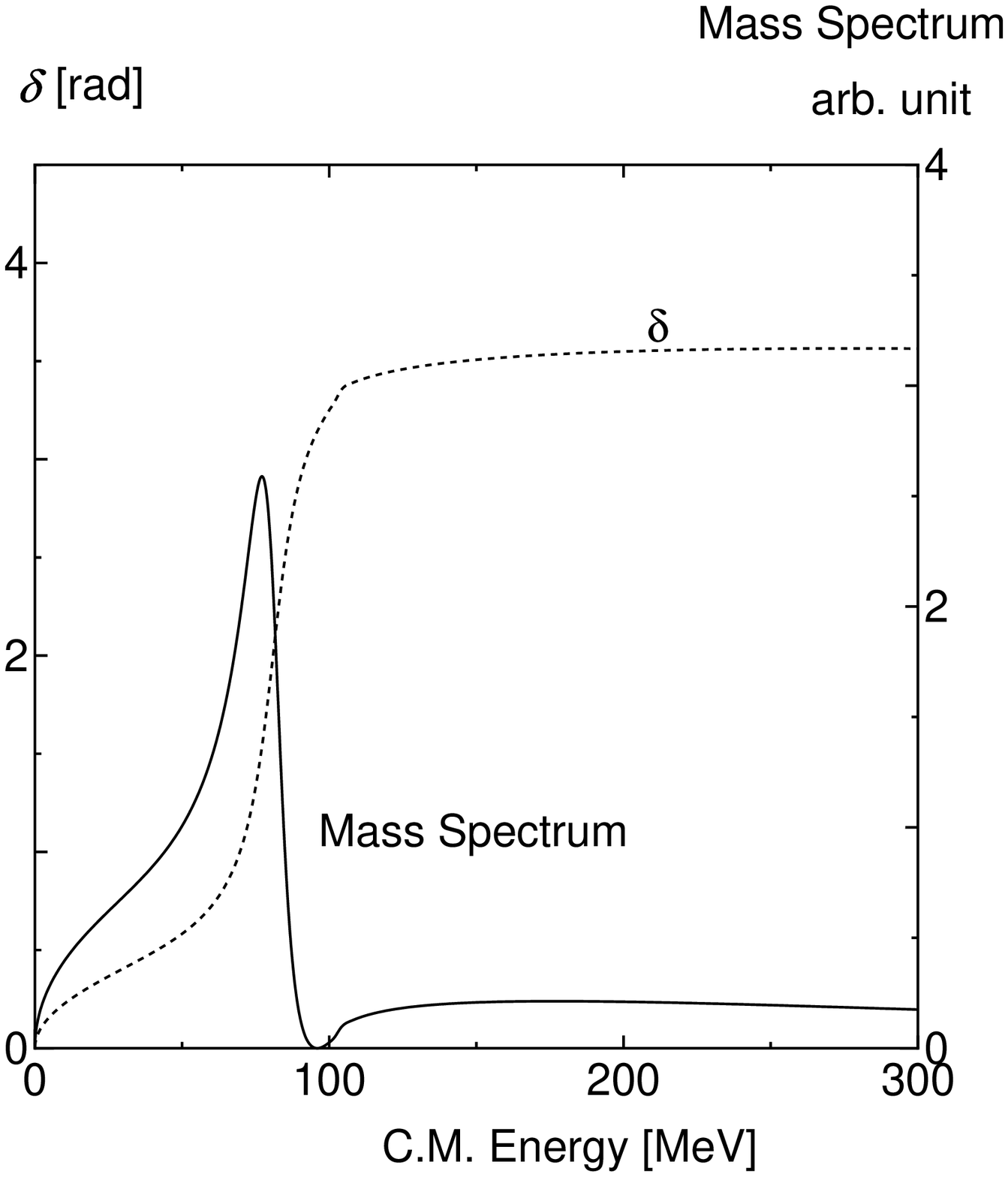}
\includegraphics[height=3.7in]{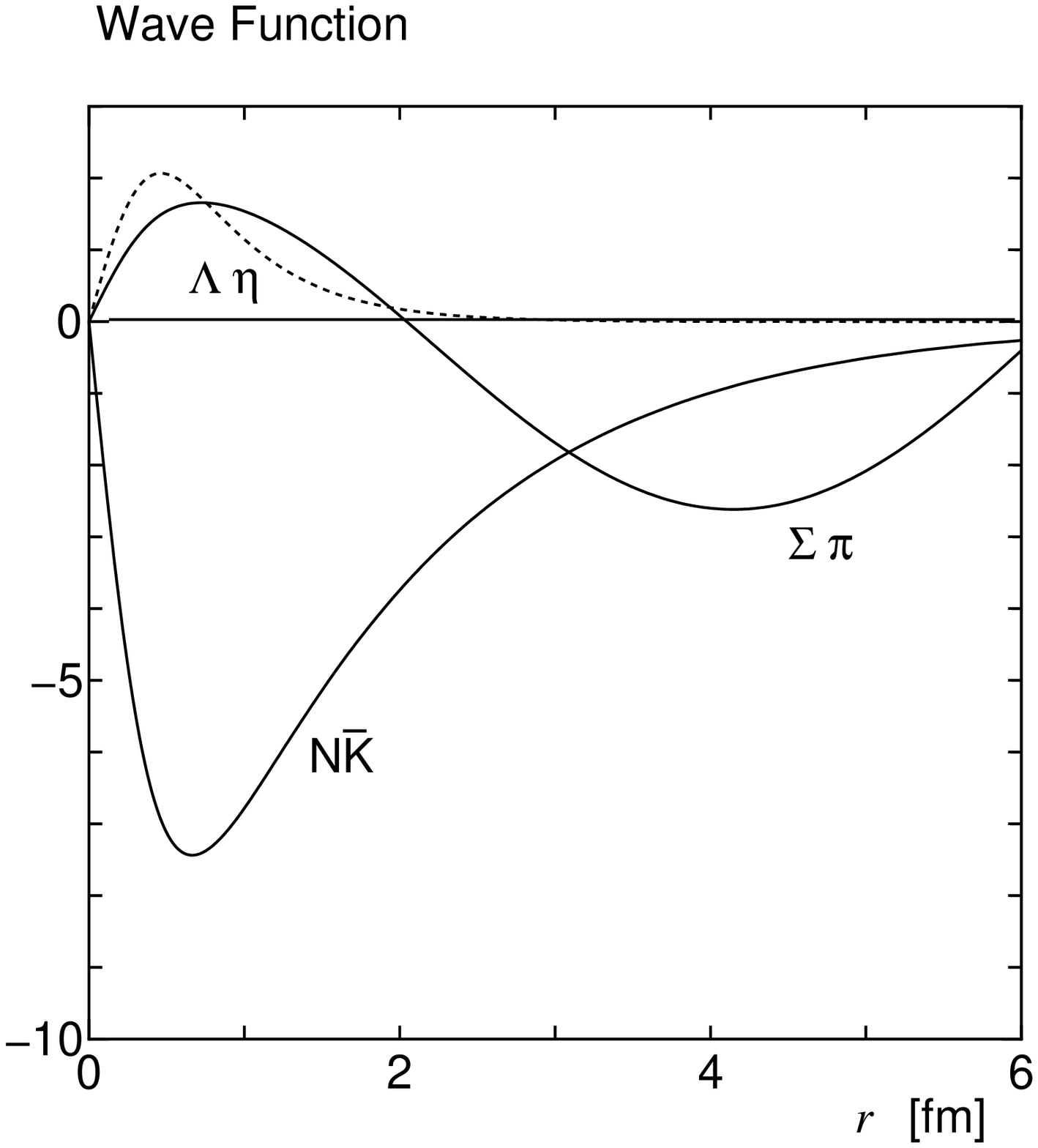}
\label{fg2}
\end{figure} 

In the Oset-type calculation, the energy dependent potential is employed. To see the effect of the energy dependence, 
we have performed a similar calculation with an energy-independent potential (parameter (2) in Table IV). 
Since the attraction between baryon and meson without the energy dependence becomes slightly weaker, 
we used a smaller range for the form factor 
in order to obtain a resonance around the same energy. As seen in Fig.\ 2, 
the shape of the mass spectrum becomes 
narrower and the N$\Kbar$ component of the wave function is larger in the short range.
Both of the cases, however, reproduce a resonance just below the N$\Kbar$ threshold.

The observed scattering length for the N$\Kbar$ $T=0$ channel is negative.
Since the FF-type interaction is strongly attractive in this channel, 
the resonance should come largely from the N$\Kbar$ bound state
in order to give a negative scattering length.  In fact,
the calculated scattering length  is negative.
It, however, 
seems somewhat larger 
in the amplitude than the observed one.
This suggests that the attraction should be decreased
 without changing the resonance energy somehow.

When a slightly long-range form factor is employed, the interaction becomes weaker.
We take $a=0.46$ fm, which corresponds to the size parameter $b=0.56$ fm in the quark model,
with the same strength $V_0$ as before (parameter (3) in Table IV).
Then, it becomes difficult 
to reproduce the resonance of $\Lambda$(1405) as seen in Fig.\ 3. 
The amplitude of the N$\Kbar$ channel 
and scattering length become very large because the resonance occurs almost on the  N$\Kbar$
 threshold.
 \medskip

\begin{figure}
\caption{Mass spectrum, phase shift and wave functions of $\Sigma \pi$ scattering
by the FF-type potential with the parameter set (3) in Table IV. 
No coupling with the q$^3$ state.} 

\includegraphics[height=4.0in]{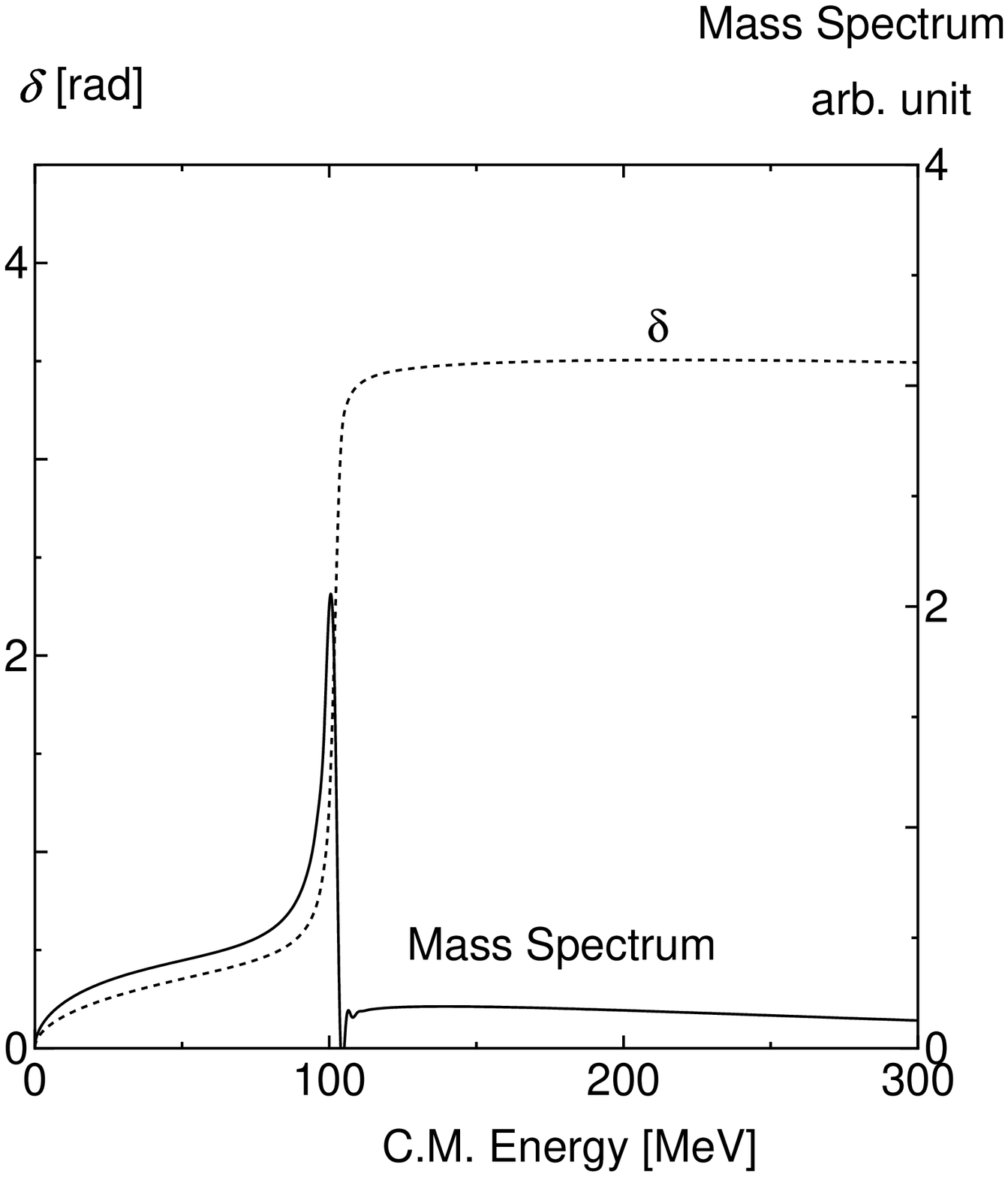}
\includegraphics[height=3.7in]{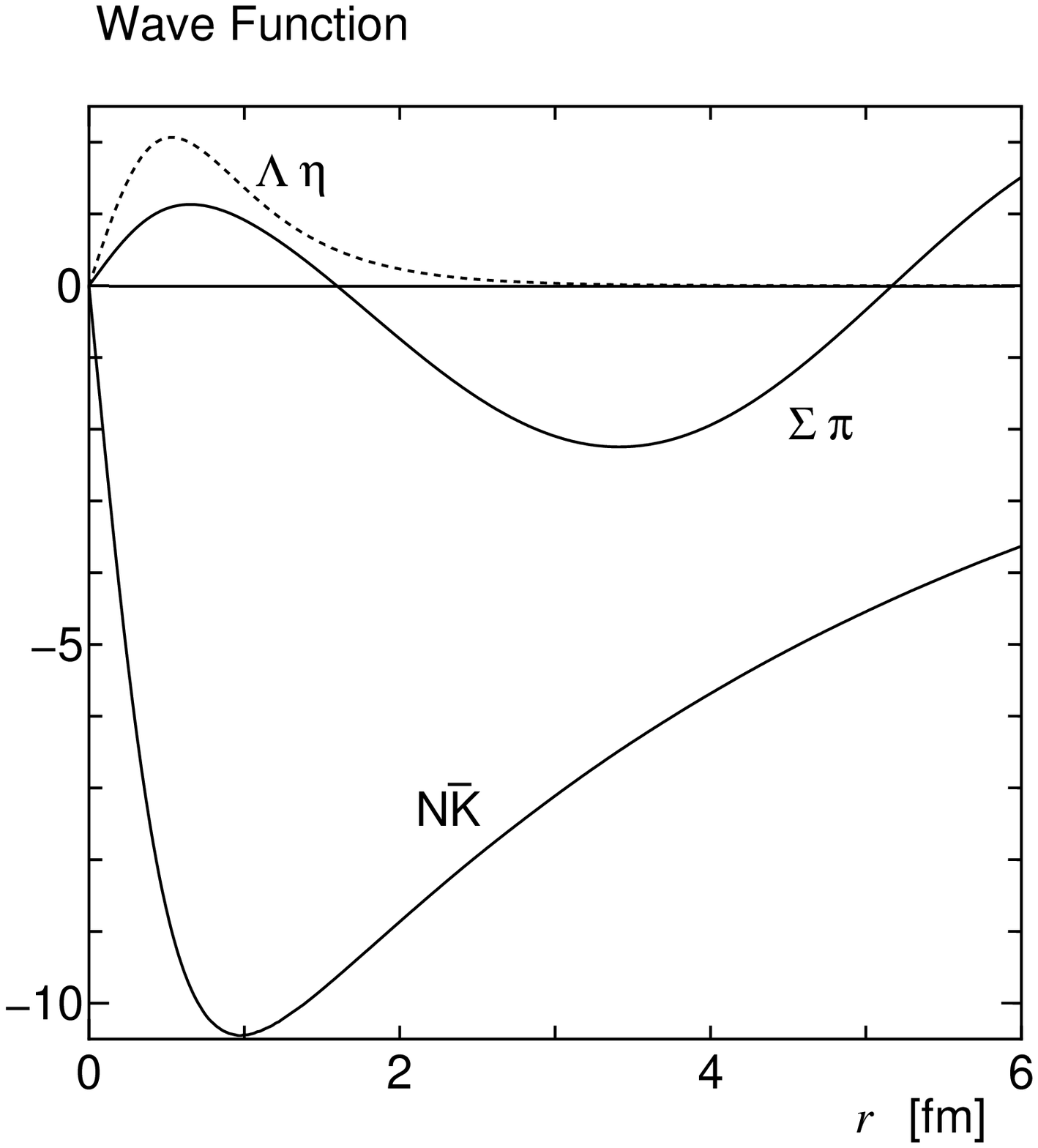}
\label{fg3}
\end{figure}

\begin{figure}
\caption{Mass spectrum, phase shift and wave functions of $\Sigma \pi$ scattering 
by the FF-type potential with the parameters (4) in Table IV.
The coupling with the state q$^3$ is $p^2$-type.} 
\includegraphics[height=4.0in]{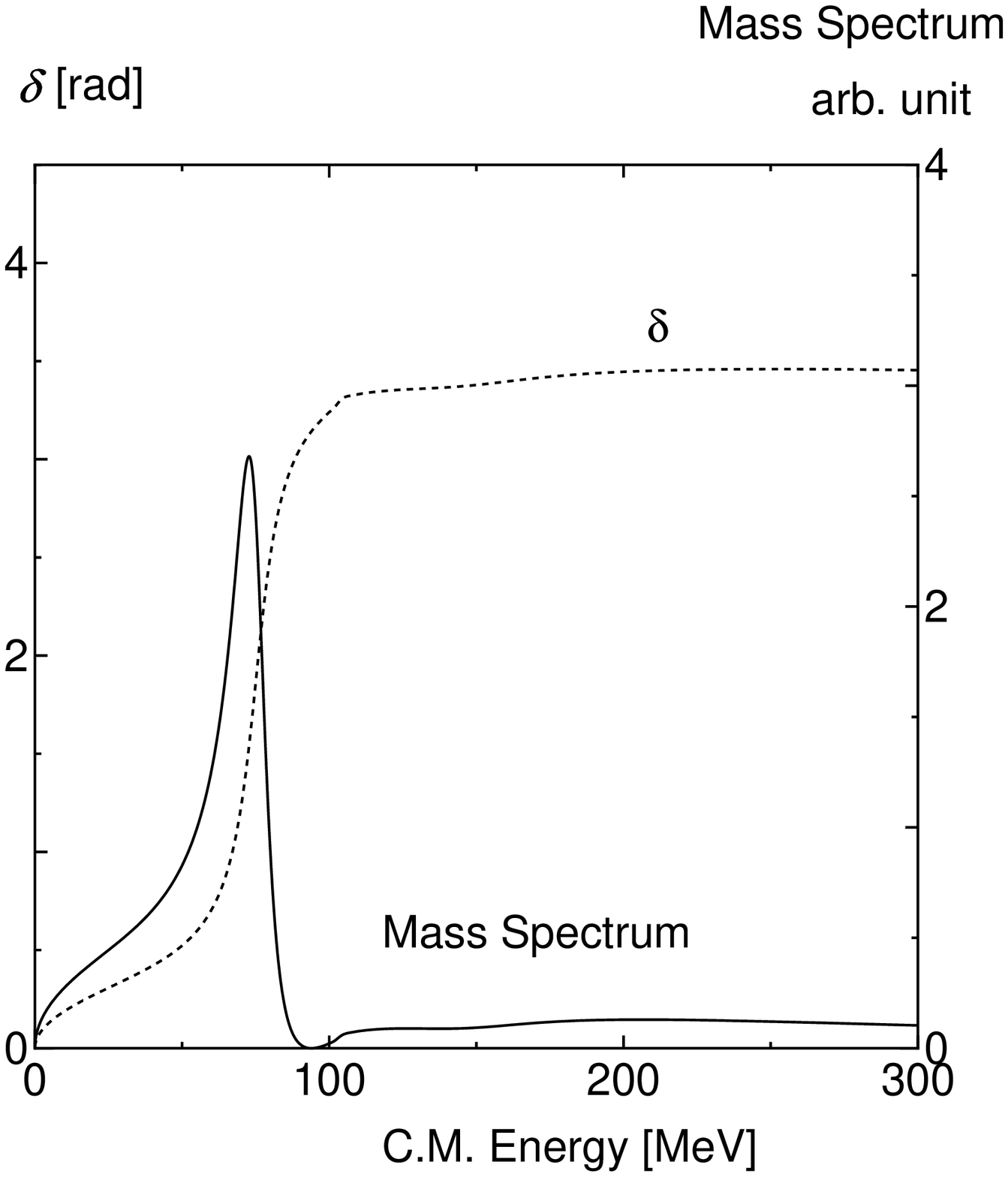}
\includegraphics[height=3.7in]{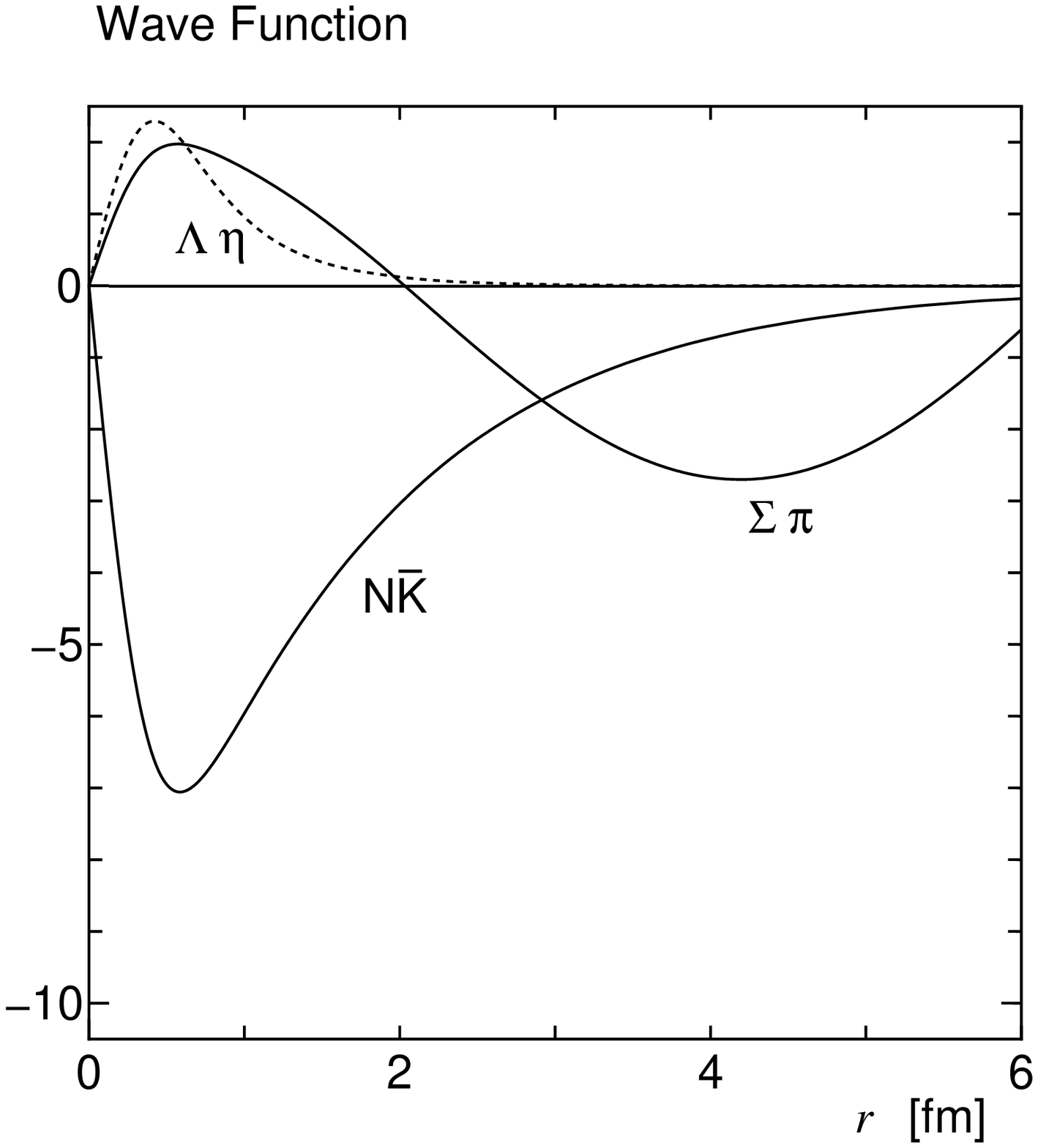}
\label{fg4}
\end{figure} 

\begin{figure}
\caption{Mass spectrum and  phase shift of $\Sigma \pi$ scattering
by the energy-dependent and independent FF-type potentials with the parameter set (4) in Table IV.
The coupling with the state q$^3$  is $p^2$-type.} 
\begin{center}
\includegraphics[height=4.0in]{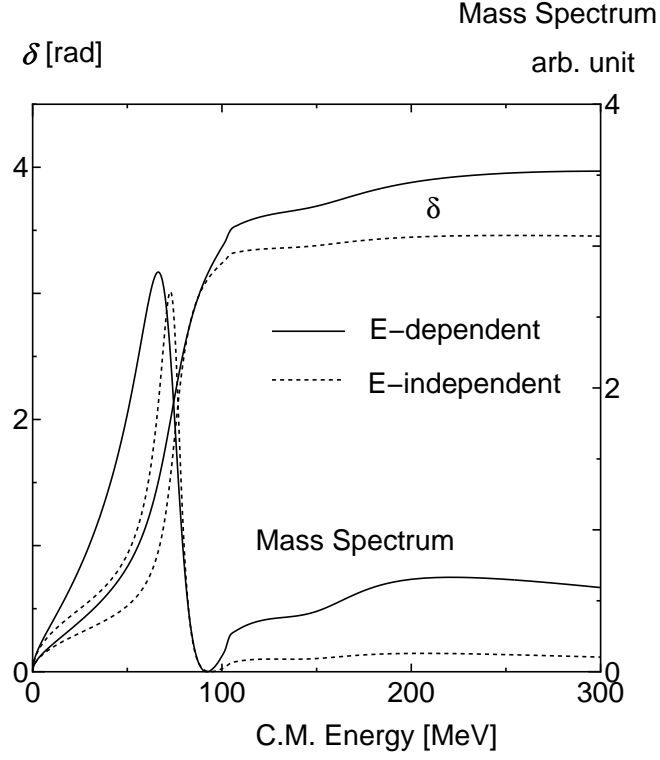}
\end{center}
\label{fg5}
\end{figure} 

\begin{figure}
\caption{Mass spectrum, phase shift and wave functions of $\Sigma \pi$ scattering
by the FF-type potential with the parameter set  (5) in Table IV.
The coupling with a state q$^3$  is 1-type.} 
\includegraphics[height=4.0in]{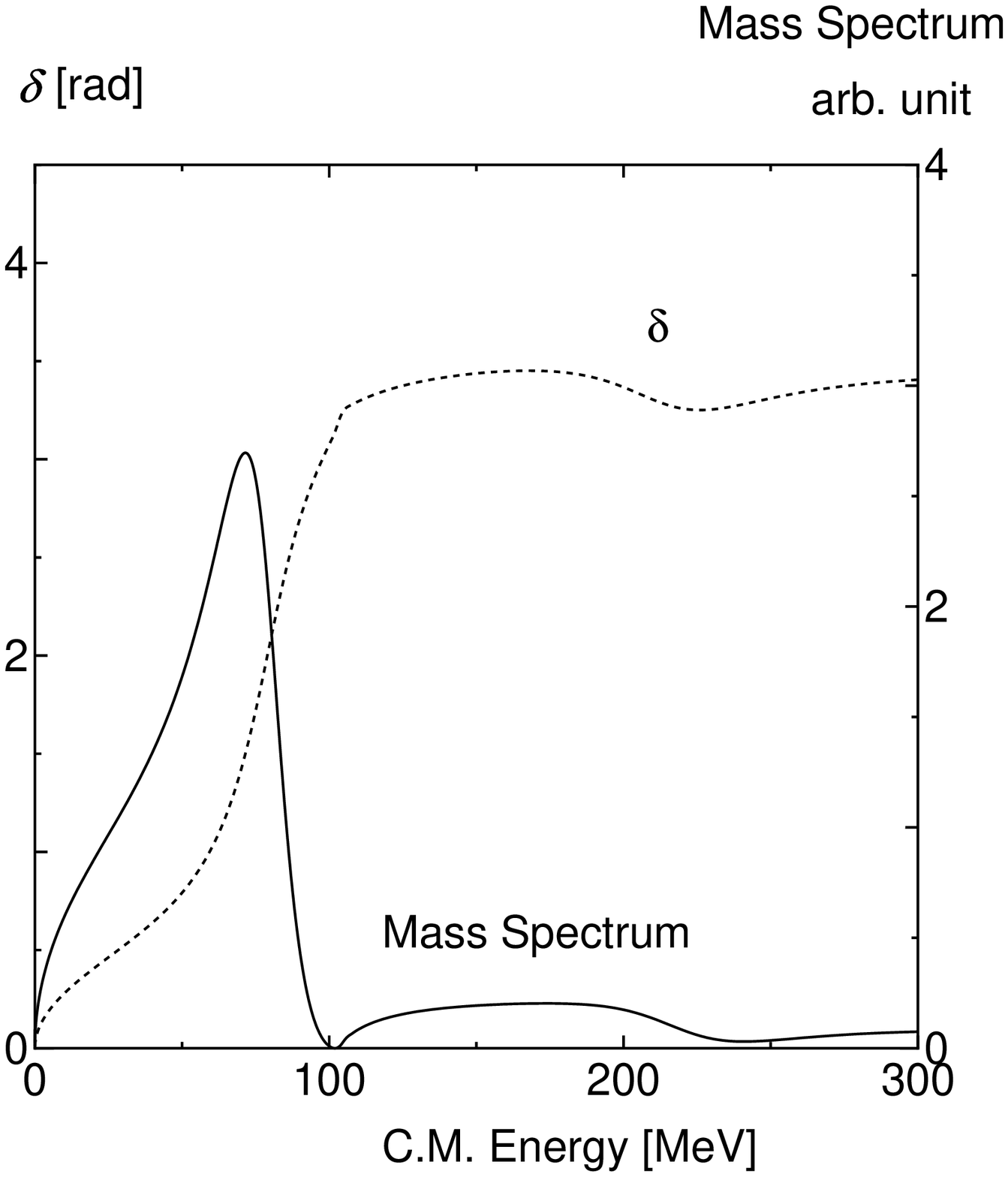}
\includegraphics[height=3.7in]{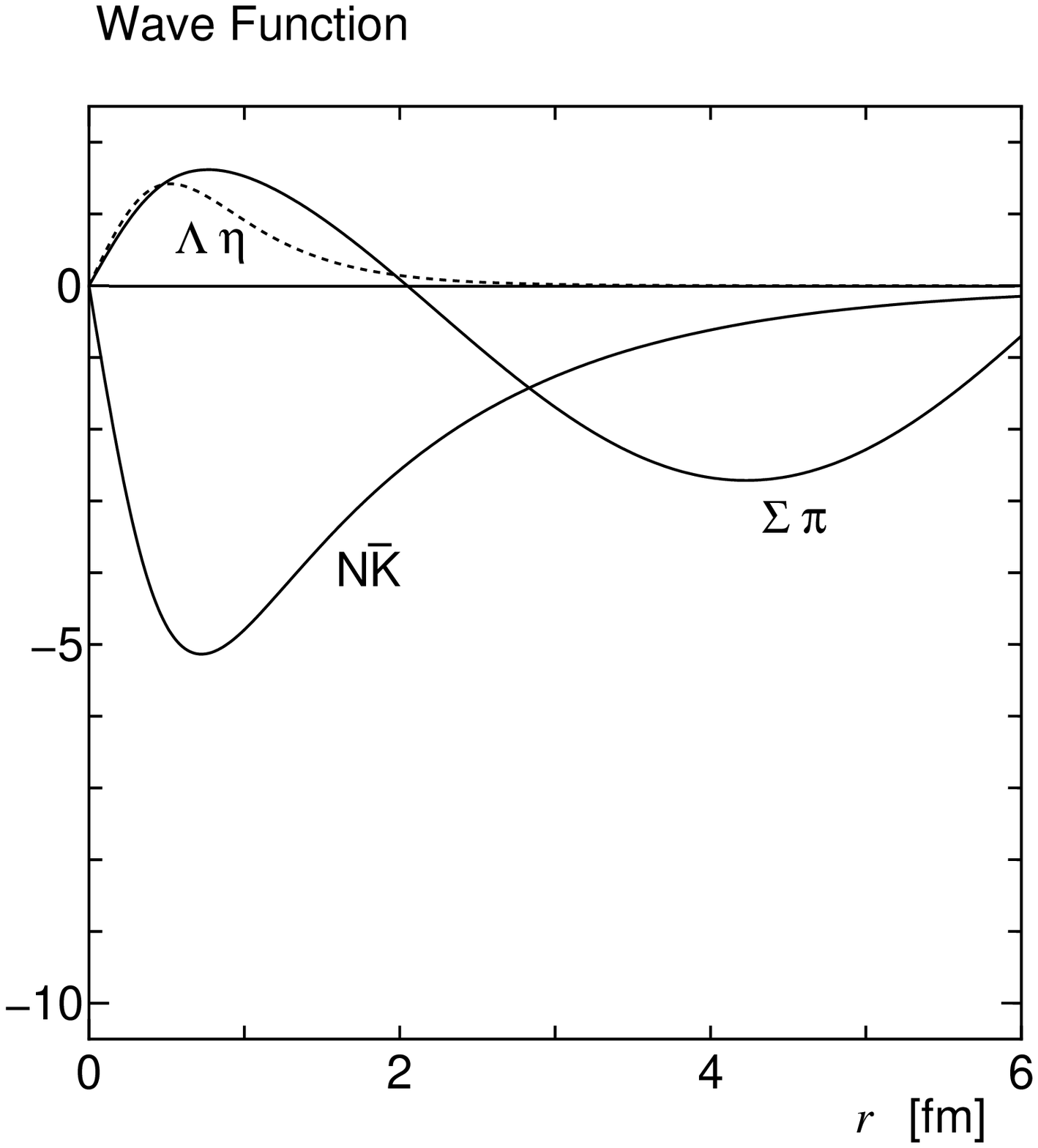}
\label{fg6}
\end{figure} 

Now let us introduce a BSEC into the system. 
In addition to the three baryon-meson channels, 
we include the flavor-singlet $P$-wave q$^3$ state as a BSEC state.
We employ 200MeV above the $\Sigma\pi$ threshold
as the mass of the $q^3$ state before the coupling is switched on.
It corresponds to 1531MeV and is slightly heavier than the mass of $\Lambda(1520)$ ($\frac{3}{2}^-$).
We take this value because 
 $\Lambda(1520)$ is considered 
to be the flavor-singlet spin partner. 
The coupling to the baryon-meson systems is much smaller here
as indicated by its small width (15.6$\pm$1.0 MeV\cite{pdg});
the unperturbed mass for  $\Lambda(1520)$ 
is considered to be close to the observed mass.
Judging from the mass spectrum of the flavor-octet excited baryons, 
the spin-orbit force is probably small also for the flavor-singlet baryons.
Thus, we choose 1531 MeV as the unperturbed q$^3$ mass for $\Lambda$(1405).

In Fig.\ 4, the mass spectrum and wave functions
with the
parameter set (4) are shown. 
Here we have employed the coupling with the BSEC which 
has $p^2$ dependence, namely, $c_1=0$, $c_2=1$ (called $p^2$-type) in eq.\ (\ref{vpq}).
The peak appears around 1404 MeV which corresponds to 
the $\Lambda(1405)$ resonance. 
Both the N$\Kbar$ and q$^3$ channels play a very important role to 
reproduce the resonance as shown in Table V. 
The q$^3$ state gains the self energy 
Re\,$\Sigma=-119.0 $ MeV 
as shown in Table VI. 
The mass spectrum is similar to the one in Fig.\ 2, which employs 
the shorter range 
form factor without the q$^3$ state. 

In order to see the effect of the energy dependence, we also 
show the results 
using the same parameters as (4) but with the energy dependence in Fig.\ 5.   As seen in the figure, the energy 
dependence gives a stronger shift of the resonance and wider shape than the one without the 
energy dependence. This is because the strength of the potential increases around the resonance.

In Fig.\ 6, we show the results of the parameter set (5), 
which employs the different structure of the 
transition potential $V_{PQ}$, namely, $c_1=1$, $c_2=0$ (called 1-type) in eq.\ (\ref{vpq}). 
As seen in the figure, the mass spectrum is similar to the one of the case (1).
It gives a smaller scattering length than that of the case (1),
but gives a wide shape of the mass spectrum. 
The self energy of the q$^3$ 
state is $\Sigma=-104.2 - 49.1~ \ii$ MeV.  

In all the above cases with a BSEC, 
we successfully produce a resonance with a large width around at 1405MeV. Here 
the attraction in the N$\Kbar$ channel plays an important role to reproduce the resonance together with the 
q$^3$ state.
The N$\Kbar$ scattering length becomes smaller reflecting the fact that the N$\Kbar$ attraction  decreases.
Introducing the q$^3$ state keeps the resonance energy at the appropriate place.
For the parameter set (5), which reproduces the observed N$\Kbar$ scattering length, 
the probability of the BSEC is roughly half of that of N$\Kbar$ at the resonance.
\medskip

So far, we have employed the FF-type potential for  the potential $V_P$, whose characteristic 
feature is the strong attractive N$\Kbar$ potential.  
In the quark cluster model, however, the channel 
dependence of the potential $V_P$ is different from the FF-type potential. 
For the parameter sets (6)-(8), we employ the CMI-type potential shown in Table II,
 which is derived from the color magnetic 
interaction with the quark exchange between q$^3$ baryon and q$\bar{\rm q}$ meson.
The strength of the CMI-type interaction is adjusted to reproduce the results of the quark cluster model.
Other parameters, such as $a$, $a_Q$, $E_Q$ and the transfer potential,
are fixed to the values of the quark cluster model \cite{Takeuchi:2007tv}.

First we use the non-relativistic Lippmann-Schwinger equation in order to reproduce the results given in 
\cite{Takeuchi:2007tv}. 
The potential parameters are (6) in Table IV and the results are shown in Table V and 
Fig.\ 7. 
The unperturbed mass of the q$^3$ state is taken to be 160 MeV above the $\Sigma\pi$ threshold.
It corresponds to 1491 MeV, following the value calculated by the q$^3$ quark model \cite{Isgur:1978xj}.
As seen in the tables and figures, we can reproduce the results without carrying out the 
complicated quark cluster model 
calculation. 

The present model is different from the original one at the point that 
the model does not include the quark degrees of freedom directly.
Our results, however, are
almost the same as the original one.
It is because the relevant channels are not affected 
from the quark Pauli-blocking effect largely\cite{Takeuchi:2002cw}.
Approximation by an energy-independent baryon-meson potential is valid for $\Lambda$(1405).

While the quark cluster model calculation is limited within the non-relativistic kinematics, 
the present approach has an advantage to employ the semirelativistic Lippmann-Schwinger equation.
Employing the CMI-type for the baryon meson potential $V_P$, 
we perform the three-channel coupled 
$\Sigma \pi$ scattering with the semirelativistic kinematics. 
We use again the two types of the coupling potential $V_{PQ}$ with the q$^3$ state, namely, 
the 1-type and the $p^2$-type, which correspond to the parameter sets (7) and 
 (8), respectively. 
As seen in Fig.\ 8, we can reproduce the resonance with a reasonable width using the 1-type coupling. 
On the other hand, the mass spectrum becomes very narrow when the $p^2$-type coupling is employed (Fig.\ 9). 
This is because the coupling with the scattering states which have large $p$ strongly pushes 
 the BSEC downwards, but 
 the decay probability is suppressed due to the $p^2$ dependence.

The difference in the types of the transfer potentials can be clearly seen
in the imaginary part of the self energy shown in Table VI.
Those with the $p^2$-type transfer potential, parameter sets (4) and (8),
give much smaller values.
This tendency is also seen in the Table V;
these parameter sets give a narrower peak and smaller imaginary part
for the N$\Kbar$ scattering length.

%%%
\begin{figure}
\caption{Mass spectrum, phase shift and wave functions of $\Sigma \pi$ scattering by a 
non-relativistic calculation 
using the CMI-type potential with the parameter set (6) in Table IV.
The coupling with a state q$^3$  is 1-type.} 
\includegraphics[height=4.0in]{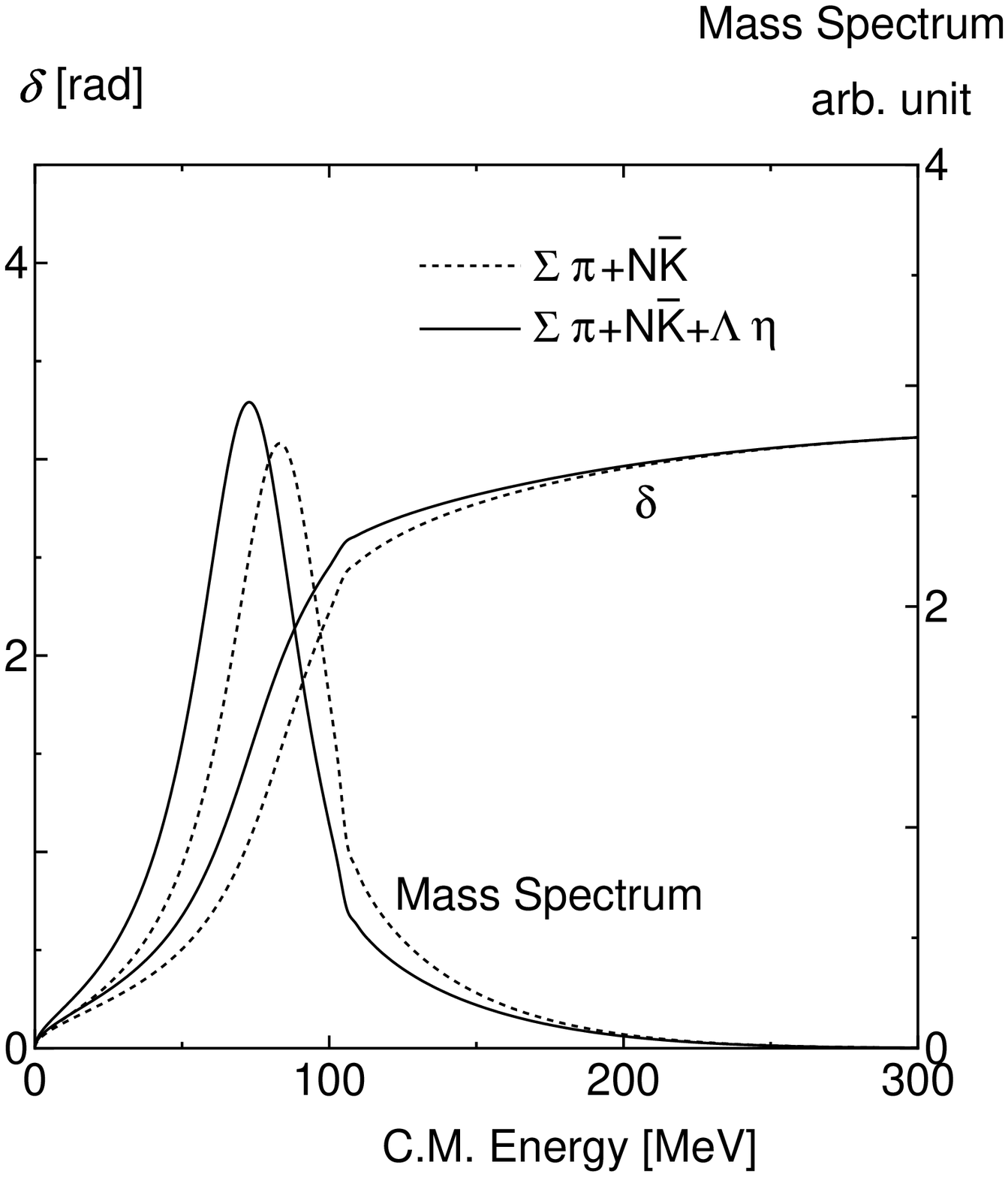}
\includegraphics[height=3.7in]{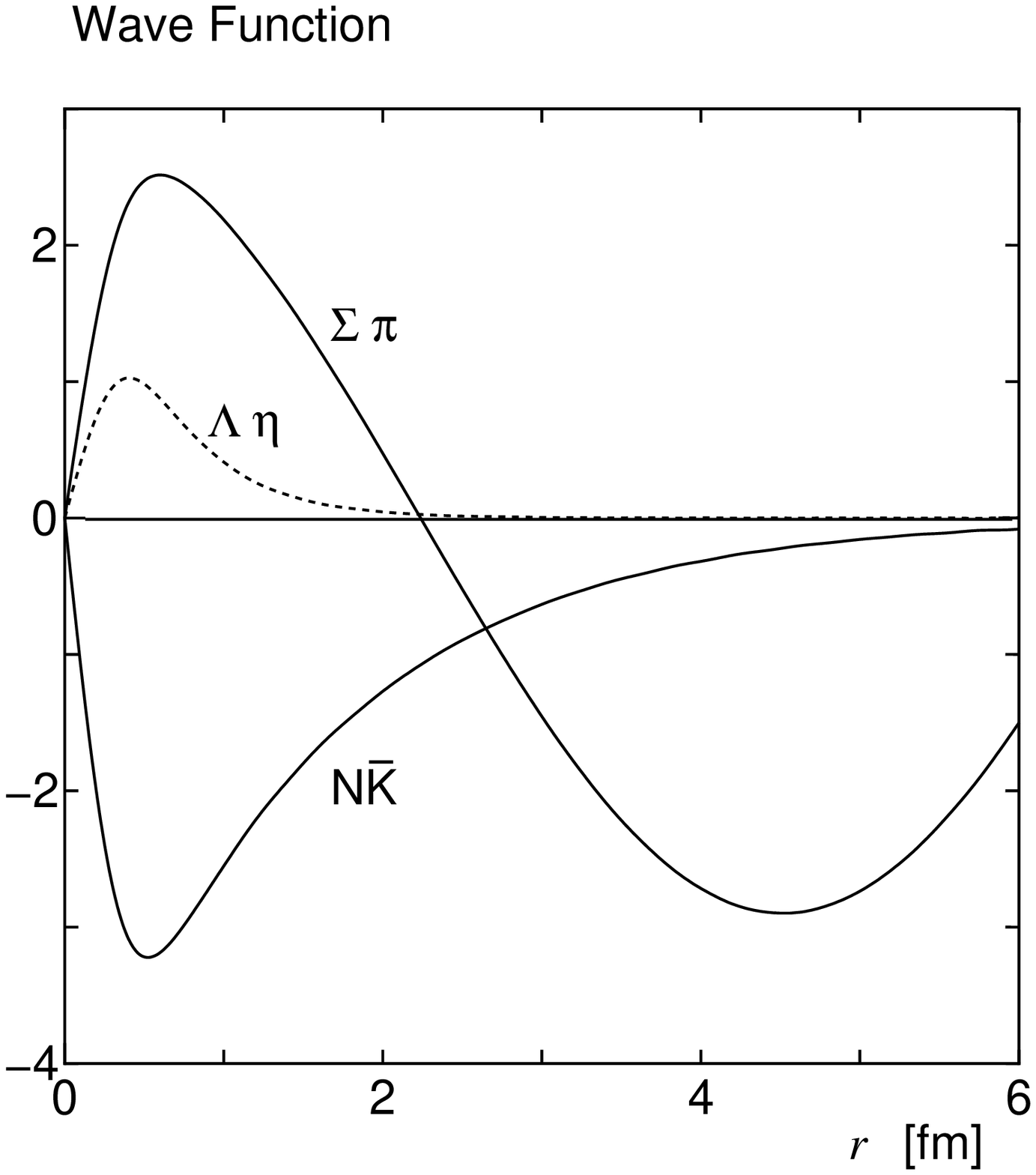}

\label{fg7}
\end{figure} 

%$\Sigma=-82.8-28.7 i$

\begin{figure}
\caption{Mass spectrum, phase shift and wave functions of $\Sigma \pi$ scattering
by the CMI-type potential with the parameter set (7) in Table IV. 
The coupling with a state q$^3$  is 1-type.} 
\includegraphics[height=4.0in]{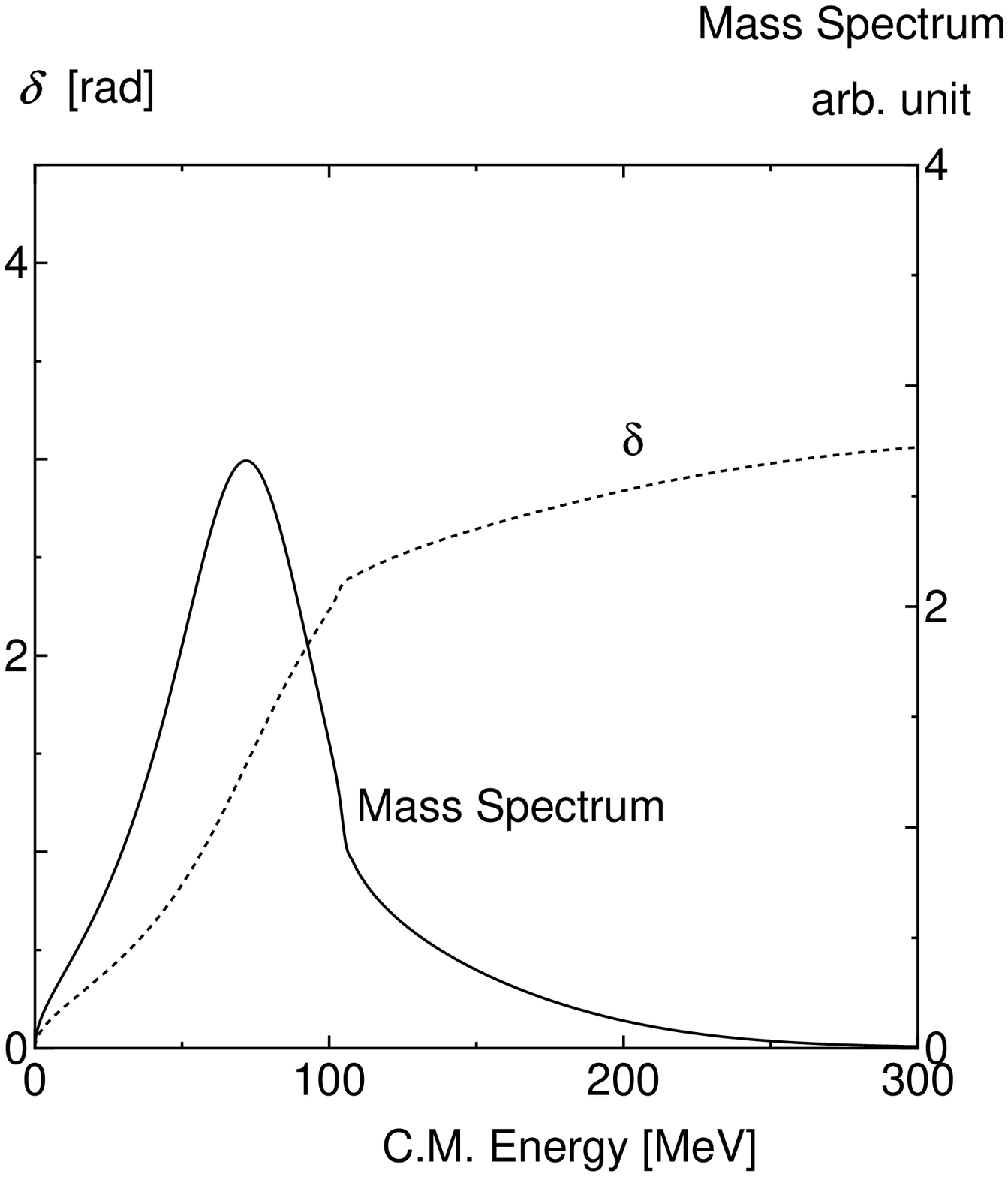}
\includegraphics[height=3.7in]{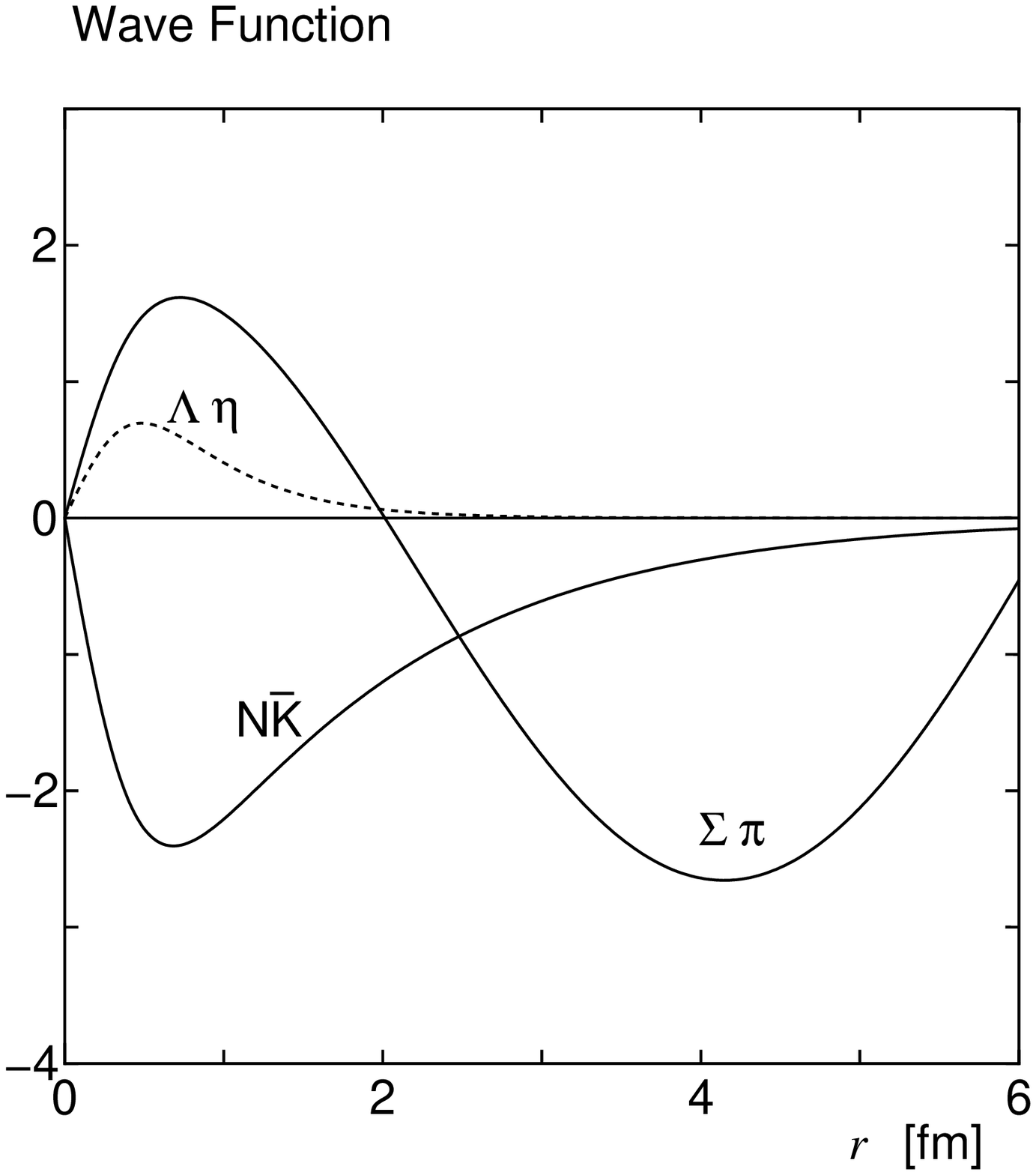}

\label{fg8}
\end{figure} 
%$\Sigma=-78.4-40.2 i$

\begin{figure}
\caption{Mass spectrum, phase shift and wave functions of $\Sigma \pi$ scattering
by the CMI-type potential with the parameters (8) in Table IV.
The coupling with a state q$^3$ is $p^2$-type.} 
\includegraphics[height=4.0in]{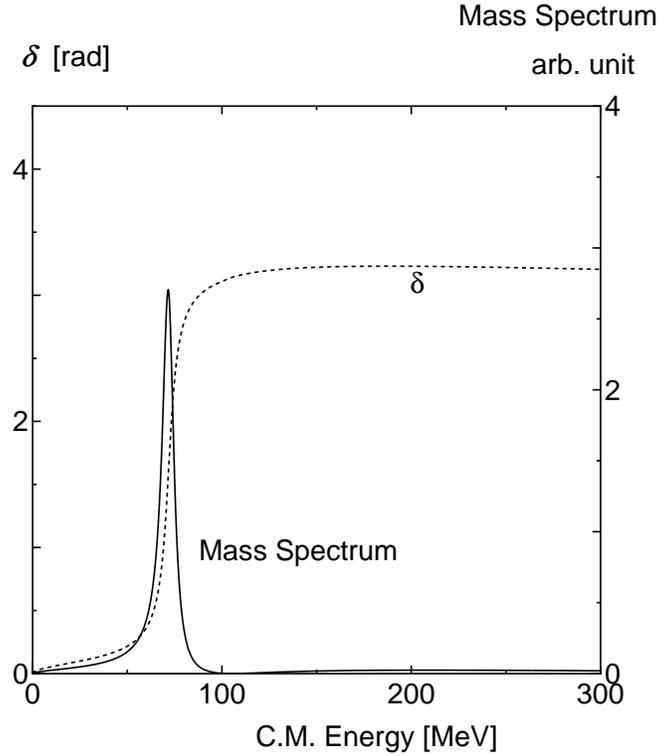}

\label{fg9}
\end{figure}

The real part of the scattering lengths given by the CMI-type models is much smaller than the observed one.
This is consistent with the larger q$^3$ contribution as seen in Table IV.
It seems that the N$\Kbar$ interaction should be more attractive than the one from the CMI-type, which may be supplied
by introducing
{\it e.g.}, the meson contribution. 
\medskip

In our parameter sets, the FF-type baryon-meson interaction with the 1-type transfer potential with the semirelativistic kinematics, the parameter set (5)
in Table IV, seems most appropriate.
As far as $\Lambda$(1405) is concerned, 
the flavor-singlet q$^3$ state is not a contradictory idea to the FF-type baryon-meson interaction.
To proceed this idea, however, 
one has to consider what this BSEC really stands for.
The FF-type interaction appears between the baryon and the meson when one assumes the 
$\rho$-meson exchange between quarks. 
But, the FF-type interaction by itself 
cannot reproduce the observed baryon or meson spectra. 
One needs some other interaction, at least, to give the hyperfine splitting.

Among other parameter sets,
the semirelativistic CMI-type baryon-meson interaction 
with the 1-type transfer potential with the semirelativistic kinematics, the parameter set (7), 
reproduces the observables well
except for the N$\Kbar$ scattering length.  
This probably shows
an additional attractive force between N$\Kbar$ should be introduced into the quark model
treatment. 
Introducing such an extra attraction is usually done to produce a realistic two-baryon 
potential in the quark cluster model.  It may comes from the $\sigma$-meson exchange
or other long-range effects.
With such an attraction, the CMI-type may give correct low energy scattering data.

\section{Summary}
We have studied the $\Lambda (1405)$ as a resonance in a coupled channel 
baryon-meson ($\Sigma \pi$-N$\Kbar$-$\Lambda \eta$) scattering. 
The meson-baryon interaction 
is taken to be a flavor-flavor type (FF-type) or 
the one whose channel dependence is the same as the quark model (CMI-type).
The former has a strong attraction in the N$\Kbar$ channel while there is 
no attraction in that channel if  the latter potential is employed.
For models which have weaker or no N$\Kbar$ attraction, we introduce 
a bound state embedded in the continuum (BSEC), 
which is considered as the flavor-singlet q$^3$ state.
The system is investigated by solving
the Lippmann-Schwinger equation with semirelativistic kinematics
in the momentum space. 

When the FF-type interaction is strong enough, 
the $\Lambda$(1405) peak is reproduced without
introducing a BSEC like the chiral unitary approach.
When the FF-type interaction is weakened by increasing the range of the form factor,
or when the CMI-type interaction is
employed, we have to introduce the BSEC, namely, the q$^3$ state
to reproduce the observed peak.

The obtained mass spectra are similar to each other.
One cannot distinguish the interactions and models 
well by the shape of the $\Lambda$(1405) peak alone.
Thus we have also investigated the relative importance of the
coupling with the N$\Kbar$ channel and the q$^3$ state
by comparing the probabilities of the N$\Kbar$
and the BSEC components at the resonance. 
The N$\Kbar$ scattering length is another clue to find the mechanism to form the resonance.

From the results of our present calculations, the followings become clarified.
The peak energy of the $\Lambda$(1405) can be reproduced by the
 above models with appropriate parameters.
The baryon-meson potential can be the FF-type or the CMI-type.
A BSEC may or may not exist for the FF-type while its existence is required
for the CMI-type.
The large width suggests that the baryon-meson potential should be 
energy-dependent or that the transfer potential between the baryon-meson system and the BSEC
does not vanish at the zero-momentum transfer.
The semirelativistic kinematics tends to give a broader peak.
The N$\Kbar$ scattering length suggests that the 
probability of the q$^3$ state is about half of that of the 
N$\Kbar$ channel at the resonance.

In our calculation, the strong FF-type interaction without a BSEC
can give an appropriate peak, but scattering length is somewhat large.
A weaker FF-type interaction 
with the q$^3$ state seems to give most appropriate results.
To proceed this idea, however, 
one has to consider what this BSEC really stands for.
One the other hand, 
the semirelativistic CMI-type interaction with the q$^3$ state
reproduces the observables well
except for the N$\Kbar$ scattering length.  
This means that 
an additional attractive force between N$\Kbar$ should be introduced to the quark model
treatment, which is also done for the two-baryon interaction
in the quark cluster model.
With such an attraction, the CMI-type may give correct low energy scattering data.
To understand the situation more clearly,
one has to derive the potentials microscopically 
from more fundamental interactions.

Let us emphasize again that a unified treatment for the resonances
should be performed to investigate the excited hadrons.
One should take into account the coupling to the quark state, or a BSEC,
in addition to the baryon-meson states, rather than to deal with 
a simple q$^3$ state or a ordinary coupled-channel scattering problem separately.

%\newpage
\vspace{1cm}

\begin{acknowledgements}
This work was supported in part by KAKENHI 
%a Grant-in-Aid for Scientific Research from JSPS 
(Nos.\ 17540264, % simizu C -2008
18042007, % takeuchi tokutei 2006 2007
and 20540281%takeuchi C 2008-
).
\end{acknowledgements}

\appendix

\section{Lippmann-Schwinger equation}
The Lippmann-Schwinger equation for $H=H_0+V$ is written as 
\begin{equation}
T=V+VG^{(0)}T, \hspace{10pt}G^{(0)}=\frac{1}{E-H_0+{\rm i}\varepsilon} .
\end{equation}
We divide a space into $P$(baryon-meson space) and $Q$(BSEC space). Assuming $V_{QQ}=0$ 
and using$P+Q=1$, we obtain the following equations.
\begin{eqnarray}
&{}&T_{PP}=V_{PP}+V_{PP}G^{(0)}_PT_{PP}+V_{PQ}G^{(0)}_QT_{QP} , \label{pp} \\
&{}&T_{QQ}=V_{QP}G^{(0)}_PT_{PQ} , \label{qq} \\
&{}&T_{QP}=V_{QP}+V_{QP}G^{(0)}_PT_{PP} , \label{qp}\\
&{}&T_{PQ}=V_{PQ}+V_{PP}G^{(0)}_PT_{PQ}+V_{PQ}G^{(0)}_QT_{QQ} . \label{pq}
\end{eqnarray}
From eq.(\ref{pp}), we obtain for $T_{PP}$,
%\[
%(1-V_{PP}G^{(0)}_P)T_{PP}=V_{PP}+V_{PQ}G^{(0)}_QT_{QP}
%\]
\begin{equation}
T_{PP}=(1-V_{PP}G^{(0)}_P)^{-1}(V_{PP}+V_{PQ}G^{(0)}_QT_{QP}).
\end{equation}
Substituting the above equation into eq.(\ref{qp}), we obtain
\begin{equation}
T_{QP}=V_{QP}+V_{QP}G^{(0)}_P(1-V_{PP}G^{(0)}_P)^{-1}(V_{PP}+V_{PQ}G^{(0)}_QT_{QP}).
\end{equation}
Introducing $G_{P}$ as  
\begin{equation}
G^{(0)}_P(1-V_{PP}G^{(0)}_P)^{-1}=(G^{(0)-1}_P-V_{PP})^{-1}=G_{P} ,
\end{equation}
we obtain
\begin{equation}
T_{QP}=V_{QP}+V_{QP}G_P(V_{PP}+V_{PQ}G^{(0)}_QT_{QP})
=(1-V_{QP}G_{P}V_{PQ}G^{(0)}_Q)^{-1}V_{QP}(1+G_PV_{PP}) .
\end{equation}
Substituting the equation into eq.(\ref{pp}), we obtain
\begin{equation}
T_{PP}=V_{PP}+V_{PP}G^{(0)}_PT_{PP}+V_{PQ}G^{(0)}_Q(1-V_{QP}G_{P}V_{PQ}G^{(0)}_Q)^{-1}
V_{QP}(1+G_PV_{PP}) .
\end{equation}
Introducing $G_{Q}$ also for $Q$ space,
\begin{equation}
G^{(0)}_Q(1-V_{QP}G_{P}V_{PQ}G^{(0)}_Q)^{-1}=(G^{(0)-1}_Q-V_{QP}G_{P}V_{PQ})^{-1}
=G_{Q} ,
\end{equation}
we obtain the following. 
\begin{equation}
T_{PP}=V_{PP}+V_{PP}G^{(0)}_PT_{PP}+V_{PQ}G_QV_{QP}(1+G_PV_{PP}) .
\end{equation}
Then we obtain for $T_{PP}$ the following result. 
\begin{equation}
T_{PP}=(1-V_{PP}G^{(0)}_P)^{-1}V_{PP}+(1-V_{PP}G^{(0)}_P)^{-1}V_{PQ}G_QV_{QP}(1+G_PV_{PP}) .
\end{equation}
Here the first term on the right hand side is the $T$-matrix $T^{(P)}$ which is 
a solution within the $P$-space.

Employing the following,
\begin{equation}
(1-V_{PP}G^{(0)}_P)^{-1}=G^{(0)-1}_PG_P=(1+V_{PP}G_P) ,
\end{equation}
$T_{PP}$ is given by
\begin{equation}
T_{PP}=T^{(P)}+(1+V_{PP}G_P)V_{PQ}G_QV_{QP}(1+G_PV_{PP}) \label{atpp} .
\end{equation}

Practical calculation can be performed by introducing the inverse of the following 
operator $M$
\begin{equation}
M=1-V_{PP}G^{(0)}_P \hspace{20pt}{\rm or}\hspace{20pt}
T^{(P)}=M^{-1}V_{PP}.
\end{equation}
Taking the transpose of the operator $M$, we obtain
$M^t$
\begin{equation}
1+V_{PP}G_P=M^{-1}, \hspace{10pt}1+G_PV_{PP}=(M^t)^{-1}.
\end{equation}

The following term which appears in the propagator $G_Q$ can be calculated using the 
same operator  $M^{-1}$ .
\begin{equation}
V_{QP}G_PV_{PQ}=V_{QP}G^{(0)}_PM^{-1}V_{PQ} .
\end{equation}

Finally we obtain the following form for the full $T$ matrix in the $P$ space.
\begin{equation}
T_{PP}=M^{-1}V_{PP}+M^{-1}V_{PQ}\frac{1}{E-E_Q-V_{QP}G^{(0)}_PM^{-1}V_{PQ}}V_{QP}(M^t)^{-1} .
\end{equation}

\section{Form factor}
We have introduced the form factor for the baryon-meson interaction $V_P$ and $V_{PQ}$. Here we make a 
brief comment on the size of the form factor and the form of the transition potential from the baryon-meson S-state 
to the negative parity baryon.

First we explain the baryon-meson and baryon vertex BBM for the $\frac{1}{2}^+$ baryons and $0^-$ meson.
Assuming that the baryon consists of 3 quarks q$^3$ whose internal wave function is given by
\begin{equation}
\phi(\xbld{\rho})\phi(\xbld{\lambda})=(\frac{1}{\sqrt{\pi}B_{\rho}})^{3/2}
\exp (-\frac{\xbld{\rho}^2}{2B_{\rho^2}})(\frac{1}{\sqrt{\pi}B_{\lambda}})^{3/2}
\exp (-\frac{\xbld{\lambda}^2}{2B_{\lambda^2}}) ,
\end{equation}
where the internal coordinates $\xbld{\rho}$ and $\xbld{\lambda}$, and the center of mass $\xbld{R}$ are
\begin{equation}
\xbld{\rho}=\xbld{r}_1-\xbld{r}_2, \hspace{10pt}\xbld{\lambda}=\frac{\xbld{r}_1+\xbld{r}_2}{2}-\xbld{r}_3,
\hspace{10pt}\xbld{R}=\frac{\xbld{r}_1+\xbld{r}_2+\xbld{r}_3}{3} .
\end{equation}
When the quark and meson vertex qqM is 1, then the form factor is given by
\begin{equation}
F(\xbld{p})=\int \phi(\xbld{\rho})^2\phi(\xbld{\lambda})^2 \exp( -i \xbld{p} \cdot \xbld{R})\exp( i \xbld{p} \cdot \xbld{r}_3)
\dd \xbld{\rho} \dd \xbld{\lambda} ,
\end{equation}
where $\xbld{p}$ is the relative momentum of the baryon and meson. 
Substituting $\xbld{r}_3=\xbld{R}-2\xbld{\lambda}/3$, we obtain
\begin{equation}
F(\xbld{p})=\exp(-\frac{B_{\lambda}^2\xbld{p}^2}{9})=\exp(-\frac{b^2\xbld{p}^2}{6}) ,
\end{equation}
where we used the size parameter of the single quark wave function $b$ given by
\begin{equation}
B_{\lambda}=\sqrt{\frac{3}{2}}b .
\end{equation}

Next we explain the form factor for the $V_{PQ}$. The internal wave function of the negative parity baryon is 
given by
\begin{equation} 
\phi(\xbld{\rho})\phi(\xbld{\lambda})=(\frac{1}{\sqrt{\pi}B_{\rho}})^{3/2} 
\exp (-\frac{\xbld{\rho}^2}{2B_{\rho^2}})
(\frac{1}{\sqrt{\pi}B_{\lambda}})^{3/2}\frac{\sqrt{2}\xbld{\lambda}}{B_{\lambda}}\exp (-\frac{\xbld{\lambda}^2}{2B_{\lambda^2}}) ,
\end{equation}
where the wave function $\phi(\xbld{\lambda})$ is taken to be  P-wave.
Taking the quark and meson vertex qqM to be the usual vertex as
\begin{equation}
\xbld{\sigma} \cdot (\xbld{p}-\frac{m}{m_q}\xbld{p}_q) ,
\end{equation}
where $\xbld{p}_q$ is the quark momentum and $\xbld{\sigma}$ is the Pauli spin matrix for quarks. Then we find that 
the baryon meson vertex contains two terms, namely, 1 and $p^2$ terms.
The form factor becomes the same as the S-wave case, and we have the following two types of the 
$V_{PQ}$ interaction.
\begin{equation}
\exp(-\frac{B_{\lambda}^2\xbld{p}^2}{9}), \hspace{20pt}
p^2\exp(-\frac{B_{\lambda}^2\xbld{p}^2}{9}) .
\end{equation}
Therefore we have taken the following form for the transition potential $V_{PQ}$
\begin{equation}
V_{PQ}(p)=V_0^{PQ}\{c_1+c_2(a_Qp)^2\}\exp (-\frac{a_Q^2p^2}{4}), \hspace{10pt}a_Q=\sqrt{\frac{2}{3}}b .
\end{equation}

\end{document}